\documentclass[epj]{svjour}
\usepackage{amsfonts}
\usepackage{epsfig}
\usepackage{rotating}

\begin{document} 

\title{Use of harmonic inversion techniques in the periodic orbit 
quantization of integrable systems}

\author{Kirsten Weibert \and J\"org Main \and G\"unter Wunner}
\institute{Institut f\"ur Theoretische Physik und Synergetik,
Universit\"at Stuttgart, D-70550 Stuttgart, Germany}

\date{\today}
%

\abstract{
Harmonic inversion has already been proven to be a powerful tool for the
analysis of quantum spectra and the periodic orbit orbit quantization 
of chaotic systems.
The harmonic inversion technique circumvents the convergence problems
of the periodic orbit sum and the uncertainty principle of the 
usual Fourier analysis, thus yielding results of high resolution
and high precision.
Based on the close analogy between periodic orbit trace formulae for
regular and chaotic systems the technique is generalized in this paper 
for the semiclassical quantization of integrable systems.
Thus, harmonic inversion is shown to be a universal tool which can be 
applied to a wide range of physical systems.
The method is further generalized in two directions:
Firstly, the periodic orbit quantization will be extended to include 
higher order $\hbar$ corrections to the periodic orbit sum.
Secondly, the use of cross-correlated periodic orbit sums allows us 
to significantly reduce the required number of orbits for semiclassical 
quantization, i.e., to improve the efficiency of the semiclassical method.
As a representative of regular systems, we choose the circle billiard,
whose periodic orbits and quantum eigenvalues can easily be obtained.
\PACS{{03.65.Sq}{Semiclassical theories and applications}
     } 
} 

\maketitle

\section{Introduction}
\label{intro}
A question of fundamental interest for systems with both regular and 
chaotic dynamics is how quantum mechanical eigenvalues can be obtained 
by quantization of classical orbits.
The EBK torus quantization method of Einstein, Brillouin, and Keller 
\cite{Ein17,Bri26,Kel58} is restricted to integrable systems, i.e.,
the method cannot be generalized to systems with an underlying chaotic
or mixed regular-chaotic dynamics \cite{Ein17}.
Furthermore, EBK quantization requires the knowledge of all the constants 
of motion, which are not normally given in explicit form, and therefore 
practical EBK quantization based on the direct or indirect numerical 
construction of the constants of motion turns out to be a formidable 
task \cite{Per77}.
As an alternative, EBK quantization was recast as a sum over all periodic 
orbits of a given topology on respective tori by Berry and Tabor \cite{Ber76}.
In contrast to EBK-quantization, {\em periodic orbit theory} can be applied 
to systems with more general classical dynamics:
Gutzwiller's trace formula \cite{Gut67,Gut90} for chaotic systems and 
the corresponding Berry-Tabor formula for regular systems \cite{Ber76} 
give the semiclassical approximation for the density of states as a sum 
over the periodic orbits of the underlying classical system. 
However, a fundamental problem of these periodic orbit sums is that they 
usually do not converge, or if they do, the convergence is extremely slow.
During recent years, various techniques have been developed to overcome 
this problem.
Most of them are especially designed for chaotic systems 
\cite{Cvi89,Aur92,Ber90} and cannot be applied to systems with regular 
or mixed regular-chaotic dynamics, or they depend on special properties of 
the system such as the existence of a symbolic dynamics.
They are therefore restricted to a relatively small number of physical 
systems.
It would be desirable to have a method at hand which is universal in the
sense that it is applicable for all types of underlying classical dynamics.

Recently, a method for periodic orbit quantization, based on harmonic 
inversion of a semiclassical signal has been developed and successfully
applied to classically chaotic systems \cite{Mai97b,Mai98,Mai99a}.
The aim of the present paper is to demonstrate that this technique is 
equally powerful in reproducing the spectra of regular systems.
The semiclassical quantization of integrable and chaotic systems on an equal
footing will be the basis for applications to systems with even
more general, i.e., mixed regular-chaotic dynamics \cite{Mai99d}.
Furthermore, the harmonic inversion technique is generalized in two directions:
Firstly, the periodic orbit quantization will be extended to include 
higher order $\hbar$ corrections \cite{Mai98c}, and, secondly, the use 
of cross-correlated periodic orbit sums \cite{Mai99c,Mai99b,Hor00} allows 
us to significantly reduce the required number of orbits for semiclassical 
quantization, i.e., to improve the efficiency of the semiclassical method.
As a representative of regular systems, we choose the circle billiard 
whose periodic orbits and quantum eigenvalues can easily be obtained.
The paper is organized as follows:

In Section \ref{POT} we give a brief overview over the periodic orbit
theory for integrable systems, especially the Berry-Tabor formula, which
is the analogue for integrable systems to Gutzwiller's trace formula for
chaotic systems. 
We then calculate the explicit expression for the density of states of
the circle billiard from the Berry-Tabor formula.
The equations are generalized in two directions, firstly, to the density
of states weighted with the diagonal matrix elements of one or more
given operators \cite{Mai99c}, and, secondly, to include higher order 
$\hbar$ corrections in the periodic orbit sum \cite{Mai98c}.

The high precision analysis of quantum spectra and the method for the
analytic continuation of non-convergent periodic orbit sums applied in
this paper are based on the {\em harmonic inversion} of time signals.
In Section \ref{filter} we briefly introduce harmonic inversion by
filter-diagonalization \cite{Wal95,Man97}.
We also discuss an extension of the filter-diagonalization method to 
cross-correlation functions \cite{Wal95,Nar97,Man98}, which can be
used to extract semiclassical eigenvalues and matrix elements from
cross-correlated periodic orbit sums with a significantly reduced set
of periodic orbits \cite{Mai99b}.

Harmonic inversion circumvents the uncertainty principle of the conventional
Fourier transform and can be used for the high precision analysis of quantum 
spectra \cite{Mai99a,Mai97a}.
In Section \ref{quantum} the method will be applied to the quantum spectra
of the circle billiard.
The analysis will verify the validity of the Berry-Tabor formula and its
generalization to spectra weighted with diagonal matrix elements discussed
in Section \ref{matrixel}.
Furthermore, harmonic inversion will be applied to determine the 
higher order $\hbar$ contributions to the periodic orbit sum. 
The Gutzwiller and the Berry-Tabor formula are only the leading order
contributions of an expansion of the density of states in terms of $\hbar$ 
and therefore only yield semiclassical approximations to the eigenvalues. 
By analyzing the difference spectrum between exact and semiclassical 
eigenvalues, first order $\hbar$ corrections to the periodic orbit sum 
can be determined, as we will demonstrate in Section \ref{hbar2}.
The results are compared with the analytic expressions for the $\hbar$ 
expansion of the periodic orbit sum given in Section \ref{hbarsec1}.

In Section \ref{poq}, we turn to the periodic orbit quantization of 
integrable systems.
Firstly, we show how in general the problem of extracting semiclassical 
eigenvalues from periodic orbit sums can be reformulated as a harmonic 
inversion problem: 
A semiclassical signal is constructed from the periodic orbit sum, the 
analysis of which yields the semiclassical eigenvalues of the system. 
The general formulae are then applied to the circle billiard and the
results are compared to the exact quantum and the EBK eigenvalues.
In Section \ref{hbar1} it is demonstrated how the accuracy of the 
semiclassical eigenvalues can be significantly improved with the help 
of higher order $\hbar$ corrections to the periodic orbit sum.

In Section \ref{crosscorr} we address the question of how to improve the
efficiency of the semiclassical quantization method, i.e., how to extract
the same number of eigenvalues with a reduced set of periodic orbits,
which is important especially when the orbits must be searched numerically.
This is achieved by constructing cross-correlated periodic orbit sums
as introduced in Section \ref{matrixel} which are then harmonically inverted
with the generalized filter-diagonalization method of Section \ref{hi2}.
The efficiency of the method will be discussed for various sets of operators
and various sizes of the cross-correlation matrix.
It is also possible to include higher order $\hbar$ corrections in the
cross-correlation signal which will allow us to calculate $\hbar$ 
corrections even for nearly degenerate states.

Some concluding remarks are given in Section \ref{conclusion}.

\section{Periodic orbit theory for integrable systems}
\label{POT}
\subsection{EBK quantization and Berry-Tabor formula}
\label{Berry-Tabor}
Integrable systems are characterized by the property that their dynamics
can be expressed in action-angle variables. 
The action variables, which are defined on certain ``irreducible''
paths, are constants of motion.
In the $2n$-dimen\-sion\-al phase space, the motion of an integrable system 
is restricted to $n$-dimensional tori, which are given by the
values of the action variables.

A well-established method for the semiclassical quantization of integrable
systems is the EBK torus quantization scheme, which was developed by
Einstein, Brillouin and Keller \cite{Ein17,Bri26,Kel58}. 
In the EBK theory, the energy eigenvalues of the system are directly
associated with certain classical tori.
These tori are defined by the EBK conditions, which select special sets 
from all possible values of the action variables of the system. 
Each such set corresponds to a quantum mechanical eigenstate of the system.
The tori selected by the EBK conditions are usually not rational, i.e.,
the orbits on these tori are usually not periodic.

For many physical systems the application of the EBK quantization scheme 
is a nontrivial task.
Especially for non-separable or near-integrable systems the irreducible
paths are difficult to find.
Most importantly, as already discussed by Einstein \cite{Ein17}
the torus quantization scheme cannot be extended to chaotic systems.
For chaotic systems, Gutzwiller derived a semiclassical expression for 
the density of states in terms of the periodic orbits of the corresponding
classical system:
The semiclassical density of states consists of a smooth background and an 
oscillating part
\begin{equation}
\rho(E)=\rho_0(E)+\rho^{\rm osc}(E)
\end{equation}
with
\begin{equation}
\label{Gutz}
 \rho^{\rm osc}(E) = {1\over\pi\hbar}\sum_{\rm po}
 {T_{\rm po}\over r|\det (M_{\rm po}-{\bf 1})|^{1/2}}
 \cos\left({S_{\rm po}\over\hbar} - \mu_{\rm po}{\pi\over 2}\right) \; .
\end{equation}
The sum runs over all periodic orbits (po) of the system,
including multiple traversals.
Here, $T$ and $S$ are the period and the action of the orbit,
$M$ and $\mu$ are the Monodromy matrix and the Maslov index, 
and the repetition number $r$ counts the traversals of the 
underlying primitive orbit.

For integrable systems an analogous formula for the density of states 
in terms of a smooth part and oscillating periodic orbit contributions
was derived by Berry and Tabor \cite{Ber76}.
While in chaotic systems the periodic orbits are isolated, the periodic 
orbits of integrable systems are all those orbits lying on rational tori -- 
i.e., tori on which the frequencies of the motion are commensurable --
and thus are non-isolated.
The Berry-Tabor formula gives the density of states in terms of the 
rational tori:
\begin{equation}
\rho(E)=\rho_0(E)+\rho^{\rm osc}(E)
\end{equation}
with
\begin{equation}
\label{BerTabForm}
 \rho^{\rm osc}(E) = {2\over\hbar^{{1\over 2}(n+1)}}
 \sum_{\bf M}
 {\cos(S_{\bf M}/\hbar-{1\over 2}\pi{\mbox{\boldmath $\alpha$}}\cdot
 {\bf M}+{1\over 4}\pi\beta_{\bf M}) \over
 |{\bf M}|^{{1\over 2}(n-1)}\  |{\mbox{\boldmath $\omega$}}_{\bf M}| \ 
 |K({\bf I}_{\bf M})|^{1\over 2}} \; .
\end{equation}
The sum runs over all rational tori at energy $E$, characterized by the 
frequency ratios given by the ray of integer numbers ${\bf M}$.
The sum includes cases where the $M_i$ are not relatively prime, 
${\bf M} = r{\mbox {\boldmath $\mu$}}$, which
corresponds to multiple traversals of the primitive periodic orbits 
on the torus characterized by {\boldmath $\mu$}.
Here, $n$ is the dimension of the system, ${\bf I}_{\bf M}$ and 
{\boldmath $\omega$}$_{\bf M}$ are the values of the action variables
and the frequencies on the torus, $S_{\bf M}$ is the action of the 
periodic orbits on the torus, and $K$ is the scalar curvature
of the energy contour.
The components of {\boldmath $\alpha$} are the 
Maslov indices of the irreducible paths on which the action variables are 
defined, and the phase $\beta$ is obtained from the second derivative
matrix of the action variables in terms of the coordinates.
In contrast to the EBK torus quantization, there is no direct relation 
between the eigenvalues of the system and the tori which enter the 
Berry-Tabor formula.

Both the EBK torus quantization and the Berry-Tabor
formula are semiclassical theories delivering lowest order 
$\hbar$ approximations to the exact quantum eigenvalues.
In general, the results of the two approaches can only be
expected to be the same in lowest order of $\hbar$ but
not necessarily beyond.
However, it was shown in Ref.~\cite{Rei96} that 
for the circle billiard, which will be discussed in the following
sections,  the two approaches are in fact equivalent
and should yield exactly the same results.

Eq.~(\ref{BerTabForm}) can be simplified for the special case $n=2$, i.e.,
for two-dimensional regular systems \cite{Ull96}:
\begin{equation}
\label{rhoUllmo}
 \rho^{\rm osc}(E) = {1\over\pi\hbar^{3/2}}
\sum_{\bf M}
 {T_{\bf M}\over M_2^{3/2}|g''_E|^{1/2}} \cos\left({S_{\bf M}\over\hbar}
 -\eta_{\bf M}{\pi\over 2}-{\pi\over 4}\right) \; .
\end{equation}
The sum runs over all rational tori at energy $E$, characterized by the 
frequency ratio given by the integer numbers ${\bf M}=(M_1,M_2)$, including 
multiple traversals (i.e., cases where $M_1,M_2$ are not relatively prime).
Here, $T_{\bf M}$ is the traversal time, and $g_E$ is the function 
describing the energy surface: $H(I_1,I_2=g_E(I_1))=E$, where $I_1$ and 
$I_2$ are the action variables. The Maslov index $\eta_{\bf M}$ is 
obtained from the Maslov indices $\alpha_1$, $\alpha_2$ of the paths on 
which the action variables are defined:
\begin{equation}
\label{Maslov}
 \eta_{\bf M} = (M_1\alpha_1+M_2\alpha_2)-\Theta(g''_E) \; ,
\end{equation}
where $\Theta$ is the Heaviside step function.

The semiclassical density of states can be expressed in terms 
of the semiclassical response function $g(E)$:
\begin{equation}
 \rho(E) = -{1\over\pi}{\rm Im}\ g(E) \; .
\end{equation}
For both chaotic and regular systems the response function is of
the form 
\begin{equation}
\label{ge1}
 g(E) = g_0(E)+\sum_{\rm po}{\mathcal A}_{\rm po} 
  e^{{i\over \hbar}S_{\rm po}} \; ,
\end{equation}
where the amplitudes are given by the Gutzwiller or the Berry-Tabor
formula, respectively. 

In practical applications both the Gutzwiller formula (\ref{Gutz}) and
the Berry-Tabor formula (\ref{BerTabForm}) suffer from the property that 
the periodic orbit sums usually do not converge.
Depending on the system in question, this problem may be overcome, e.g., 
by convolution of the periodic orbit sum with a suitable averaging 
function \cite{Rei96}.
But even then the convergence will usually be slow, and a large number of 
orbits has to be included in order to obtain the semiclassical eigenvalues.
In the following sections, we will demonstrate how the convergence problem 
can be circumvented by the harmonic inversion method and the eigenvalues 
can be calculated from a relatively small number of periodic orbits.

\subsection{Application to the circle billiard}
\label{circle}
We now apply the Berry-Tabor formula to the circle billiard
as a specific separable system with two degrees of freedom.

For completeness and comparisons with the results from periodic orbit 
theory we first briefly review the quantum mechanical expressions
and the EBK quantization condition.
The exact quantum mechanical energy eigenvalues of the circle billiard 
with radius $R$ are given by the condition
\begin{equation}
\label{jl=0}
 J_m(kR) = 0 \; , \quad 
 m \in {\mathbb Z} \; , \quad
 E = {\hbar^2k^2\over 2M} \; ,
\end{equation}
where $J_m$ are the Bessel functions of integer order.
Here, $M$ denotes the mass of the particle, $E$ is the energy, 
and $k=\sqrt{2ME}/\hbar$ is the wavenumber.
The corresponding wave functions are given by
\begin{equation}
 \psi(r,\varphi) = J_m(kr) e^{im\varphi} \; .
\end{equation}
As $J_{-m}(x)=(-1)^m J_m(x)$, all energy eigenvalues belonging to nonzero
angular momentum quantum number $(m\ne 0)$ are twofold degenerate.
In the following the exact quantum mechanical results for the circle billiard 
are used as a benchmark for the development and application of semiclassical 
quantization methods for integrable systems.

The circle billiard problem in two dimensions is separable in polar 
coordinates. 
The semiclassical expressions for both EBK torus quantization and the
Berry-Tabor formula for the density of states are based on the action-angle
variables associated with the angular $\varphi$-motion and the radial 
$r$-motion \cite{Rei96,Meh99}.
The action variables are given by
\begin{eqnarray}
 I_\varphi &=& {1\over 2\pi}\oint p_\varphi\ d\varphi = L \\
 I_r &=& {1\over 2\pi}\oint p_r\ dr  \nonumber \\
 &=& {1\over\pi}\left(\sqrt{2MER^2-L^2}-
|L|\arccos{|L|\over\sqrt{2ME}R}\right) ,
\end{eqnarray}
where $E$ and $L$ are the energy and the angular momentum, respectively.
Quantization of the action variables
\begin{eqnarray}
 I_\varphi &=& \left(m+{\alpha_\varphi\over 4}\right)\hbar \; , \quad
   m \in {\mathbb Z} 
\label{Iphi}\\
 I_r &=& \left(n+{\alpha_r\over 4}\right)\hbar \; , \quad
   n = 0, 1, 2, \dots 
\label{Ir}
\end{eqnarray}
with $\alpha_\varphi=0$ and $\alpha_r=3$ for the circle billiard
yields the EBK quantization condition
\begin{equation}
\label{EBKev}
 \sqrt{(kR)^2-m^2}-|m|\arccos{|m|\over kR} = \left(n+{3\over 4}\right)\pi \; ,
\end{equation}
where $L=m\hbar$ are the angular momentum eigenvalues.

\begin{figure}
\vspace{6cm}       
\includegraphics{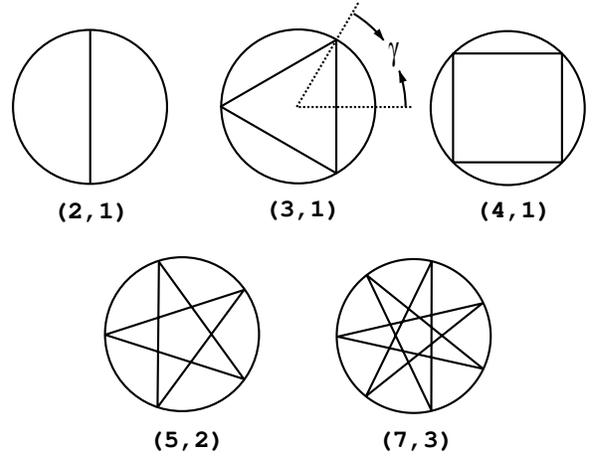}
\caption{Some examples of periodic orbits of the circle billiard.
The orbits are labeled with the numbers $(M_r,M_\varphi)$ which
correspond to the number of sides of the polygons and the number
of turns around the center. The angle $\gamma$ is given by
$\gamma=\pi M_\varphi/M_r$.
\label{fig1}
}
\end{figure}

The frequencies of the classical motion on the two-dimensional tori 
are given by
\begin{eqnarray} 
 \omega_\varphi &=& {\partial E\over \partial I_\varphi}
   = {2E \over \sqrt{2MER^2-L^2}}\arccos{|L|\over\sqrt{2ME}R}\\
 \omega_r &=& {\partial E\over \partial I_r} 
   = {2\pi E \over \sqrt{2MER^2-L^2}} \; .
\end{eqnarray}
The Berry-Tabor formula includes all tori with a rational frequency ratio, 
i.e., tori on which the orbits are periodic. 
In the case of the circle billiard, the rational tori are given by the 
condition 
\begin{equation}
 {\omega_\varphi \over \omega_r}={M_\varphi\over M_r} 
\end{equation}
with positive integers $M_r$, $M_\varphi$ and the restriction
\begin{equation}
 M_r \ge 2M_\varphi \; .
\end{equation}
The periodic orbits of the circle billiard have the form of regular polygons.
The numbers $M_r$ and $M_\varphi$ can be shown to be identical with the 
number of sides of the corresponding polygon and its number of turns around 
the center of the circle, respectively \cite{Bal72}. 
Some examples are given in Figure \ref{fig1}.
\def\baselinestretch{1}
\def\baselinestretch{2}
A pair of numbers $(M_r,M_\varphi)$ which are not relatively prime
corresponds to multiple traversals of a primitive periodic orbit.

The classical action of the periodic orbits is given by
\begin{eqnarray}
 S_{\bf M} &=& 2\pi M_\varphi I_\varphi^{({\bf M})}+2\pi M_r I_r^{({\bf M})}
 \nonumber \\
  &=& \sqrt{2ME}R\ 2 M_r\sin\left({M_\varphi\over M_r}\pi\right) \; .
\end{eqnarray}
As in all billiard systems, the action scales like
\begin{equation}
\label{scaling}
 S/ \hbar = w s \; ,
\end{equation}
here with the scaling parameter 
\begin{equation}
\label{defw}
 w \equiv \sqrt{2ME}\ R/ \hbar = kR \; 
\end{equation}
and the scaled action
\begin{equation}
\label{defs}
 s \equiv 2 M_r\sin\left({M_\varphi\over M_r}\pi\right) \; .
\end{equation}
The form of the corresponding classical trajectory is 
independent of $w$.
For the circle billiard with unit radius $R=1$, the scaling parameter $w$ 
is identical with the wavenumber $k$, and the scaled action is the length 
of the orbit.

For the semiclassical density of states, we start from the special version
of the Berry-Tabor formula presented in Eq.~(\ref{rhoUllmo}).
Using the relation 
\begin{equation}
\label{defrhow}
 \rho(w) = {dE\over dw}\ \rho(E) \; ,
\end{equation}
valid for billiard systems, we introduce the density of states depending on
the scaling parameter $w$.
Evaluating the different expressions in (\ref{rhoUllmo}) for the circle 
billiard then finally leads to 
\begin{equation}
 \rho^{\rm osc}(w) = -{1\over\pi}{\rm Im}\ g^{\rm osc}(w) \; ,
\end{equation}
with
\begin{equation}
\label{gw}
 g^{\rm osc}(w) = \sqrt{{\pi\over 2}} \sqrt w
 \sum_{\bf M}m_{\bf M}{s_{\bf M}^{3/2}\over M_r^2}
 e^{i(ws_{\bf M}-{3\over 2}M_r\pi-{\pi\over 4})} \; ,
\end{equation}
where we have used the relations $\alpha_\varphi=0$ and $\alpha_r=3$ for the
Maslov indices.
The sum runs over all pairs of positive integers ${\bf M}=(M_r,M_\varphi)$ 
with $M_r\ge 2M_\varphi$. 
The degeneracy factor 
\begin{equation}
m_{\bf M} = \left\{
\begin{array}{ll}
1 \; ; \quad & M_r = 2M_\varphi\\
2 \; ; \quad & M_r > 2M_\varphi
\end{array}\right. ,
\end{equation}
accounts for the fact that all trajectories with $M_r\ne 2M_\varphi$ 
can be traversed in two directions.

\begin{figure}
\vspace{6cm}       
\includegraphics{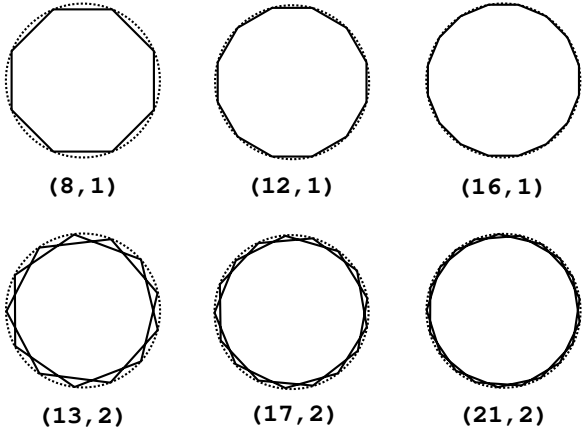}
\caption{Behavior of the periodic orbits of the circle billiard
for large $M_r$. 
The orbits are labeled with the numbers $(M_r,M_\varphi)$.
 As $M_r\to\infty$ at constant $M_\varphi$, the shape
of the orbit converges against $M_\varphi$ times the circular border 
of the billiard. 
The length of each single side of the orbits tends to zero, while the
total length of the orbit $2M_r\sin(\pi M_\varphi/M_r)R$
converges against $2\pi M_\varphi R$ ($R$: radius of the billiard). 
The circle billiard therefore possess accumulation points of
orbits at scaled actions of multiples of $2\pi$.
\label{fig2}
}
\end{figure}

Due to the rapid increase of the number of periodic orbits with growing 
action, the sum (\ref{gw}) does not converge. 
In our case, the problem is even more complicated by the fact that there 
exist accumulation points of periodic orbits at scaled actions of multiples 
of $2\pi$ (see Fig.~\ref{fig2}), which means that we cannot even include 
all periodic orbits up to a given finite action.
\def\baselinestretch{1}
\def\baselinestretch{2}
In Ref.~\cite{Rei96} the convergence problem of the sum (\ref{gw}) was
solved by averaging it with a Gaussian function.
The semiclassical eigenvalues of the circle billiard were calculated from
the periodic orbit sum by including a very large number of periodic orbits.
Our aim is to demonstrate that by using the harmonic inversion scheme, 
we can obtain eigenvalues of $w=kR$ from a relatively small number of 
periodic orbits.
We will return to this problem in Section \ref{poq}.

\subsection{Semiclassical matrix elements}
\label{matrixel}
The semiclassical trace formula for both regular and chaotic systems
can be extended to include diagonal matrix elements.
The calculation of individual semiclassical matrix elements is an 
objective in its own right.
Furthermore, the extended trace formulae will allow us to construct 
cross-correlated periodic orbit signals and thus to significantly 
reduce the required number of orbits for periodic orbit
quantization, as we will demonstrate in Section \ref{crosscorr}.
Here, we will briefly recapitulate the basic ideas and equations.

Both Gutzwiller's and Berry and Tabor's formula give the semiclassical 
response function as a sum over contributions from periodic orbits
(see Eq.~(\ref{ge1})).
The quantum mechanical response function is the trace over the Green 
function $G^+_E$,
\begin{equation}
\label{gqme}
 g_{\rm qm}(E) = \sum_n {1\over E-E_n+i0} = {\rm tr}\ G^+_E \; .
\end{equation}
As a generalization, one can consider the quantum mechanical response 
function weighted with the diagonal matrix elements of some operator 
$\hat A$, i.e.,
\begin{equation}
\label{gaqme}
   g_{A,{\rm qm}}(E) = \sum_n {\langle n|\hat A|n \rangle\over E-E_n+i0}
 = {\rm tr}\ (G^+_E\hat A) \; .
\end{equation}
The semiclassical approximation to the extended response function 
(\ref{gaqme}) is obtained by weighting the contributions of the periodic 
orbits in (\ref{ge1}) with the average $\bar A_p$ of the corresponding 
classical quantity $A({\bf q},{\bf p})$ over the periodic orbits:
\begin{equation}\label{g_A(E)}
 g_A(E) = g_{A,0}(E)+\sum_{\rm po}{\mathcal A}_{\rm po}\bar A_p 
  e^{{i\over \hbar}S_{\rm po}} \; .
\end{equation}
For chaotic systems, the average is taken over one period $T_p$ of the
isolated periodic orbit \cite{Wil87,Wil88,Eck92}:
\begin{equation}
\label{average}
 \bar A_p = {1\over T_p}\int_0^{T_p} A({\bf q}(t),{\bf p}(t))\ dt \; .
\end{equation}
For an $N$-dimensional integrable system, the quantity $A$ has to be
expressed in action-angle variables (${\bf I}$, {\boldmath $\theta$}) 
and averaged over the rational torus \cite{Meh99}:
\begin{equation}
\label{average_reg}
 \bar A_p = {1\over (2\pi)^N}\int 
 A({\bf I},{\mbox{\boldmath $\theta$}}) d^N\theta\; .
\end{equation}

Eq.~(\ref{gaqme}) can even be further generalized by introducing a second 
operator $\hat B$ and considering the quantity \cite{Mai99c}
\begin{equation}
\label{gabqm}
   g_{AB,{\rm qm}}(E)
 = \sum_n {\langle n|\hat A|n\rangle
           \langle n|\hat B|n\rangle\over E-E_n+i0} \; .
\end{equation}
If either $\hat A$ or $\hat B$ commutes with the Hamiltonian, 
Eq.~(\ref{gabqm}) can be written as a trace formula and a calculation 
similar to that in \cite{Eck92} yields the semiclassical approximation 
\begin{equation}
\label{gab}
 g_{AB}(E) = g_{AB,0}(E)+\sum_{\rm po}{\mathcal A}_{\rm po}\bar A_p\bar B_p
  e^{{i\over \hbar}S_{\rm po}} \; .
\end{equation}
Note that for general operators $\hat A$ and $\hat B$, Eq.~(\ref{gabqm}) 
cannot be written as a trace any more.
However, strong numerical evidence was provided (for both regular and chaotic 
systems) that Eq.~(\ref{gab}) is correct in general, i.e., even if neither 
operator $\hat A$ nor $\hat B$ commutes with the Hamiltonian \cite{Mai99c}.
For chaotic systems, a mathematical proof of Eq.~(\ref{gab}) is given 
in Ref.~\cite{Hor00}.
An analogous rigorous derivation for integrable systems is, to our knowledge, 
still lacking.

In Refs.~\cite{Mai99c,Hor00}, the relations (\ref{gabqm}) and (\ref{gab}) 
were generalized to products of diagonal matrix elements of more than two 
operators. 
As a further extension, we can also introduce functions of diagonal
matrix elements in the response function:
\begin{equation}
\label{gfqm}
   g_{f(A),{\rm qm}}(E)
 = \sum_n {f(\langle n|\hat A|n\rangle)\over E-E_n+i0} \; .
\end{equation}
By a Taylor expansion of the (sufficiently smooth) function $f$ and using 
the results of Refs.~\cite{Mai99c,Hor00} for multiple products of matrix 
elements, we obtain the semiclassical approximation 
\begin{equation}
\label{gf}
 g_{f(A)}(E) = g_{f(A),0}(E)+\sum_{\rm po}{\mathcal A}_{\rm po}f(\bar A_p)
  e^{{i\over \hbar}S_{\rm po}} \; .
\end{equation}
We will use the extended trace formulae in combination with 
an extension of the harmonic inversion procedure to cross-correlated
signals in order to significantly reduce the number of orbits which
have to be included in the periodic orbit sum.

The diagonal matrix elements obtained from the extended
trace formulae are semiclassical
approximations to the exact quantum matrix elements. For the circle
billiard, we can compare these values to those given by EBK theory.
According to EBK theory, the diagonal matrix element of an operator
$\hat A$ with respect to an eigenstate $|n\rangle$ is 
obtained by averaging the corresponding classical quantity $A({\bf I},{\mbox{\boldmath $\theta$}})$
over the quantized torus related to this eigenstate:
\begin{equation}
\langle n| \hat A| n\rangle = 
{1\over (2\pi)^N}\int A({\bf I}_n,{\mbox{\boldmath $\theta$}}_n) d^N\theta_n
\end{equation}
Note the difference to Eq.~(\ref{average_reg}), where the
average is taken over the rational tori.

\subsection{Higher order $\hbar$ corrections}
\label{hbarsec1}
The Berry-Tabor formula for integrable systems and Gutzwiller's trace formula
for chaotic systems are only the leading order terms of an expansion of the 
density of states in terms of $\hbar$. 
In billiard systems, the scaling parameter $w$ of the classical 
action (cf.\ Eq.~(\ref{scaling}))
is proportional to $\hbar^{-1}$ and thus plays the role of an inverse 
effective Planck constant,
\begin{equation}
\label{hbar_eff}
 w = \hbar_{\rm eff}^{-1} \; .
\end{equation}
The $\hbar$ expansion of the response function can therefore be written 
as a power series in terms of $w^{-1}$ \cite{Mai98c}:
\begin{equation}
   g^{\rm osc}(w)
 = \sum_{n=0}^\infty g_n(w)
 = \sum_{n=0}^\infty {1\over w^{n}} \sum_{\rm po} 
   {\cal A}_{\rm po}^{(n)} e^{is_{\rm po}w} .
\label{g_hbar_series}
\end{equation}
The zeroth order amplitudes ${\cal A}_{\rm po}^{(0)}$ are those of 
the Berry-Tabor or Gutzwiller formula, respectively, whereas for $n>0$, 
the amplitudes ${\cal A}_{\rm po}^{(n)}$ give the $n^{\rm th}$ order 
corrections $g_n(w)$ to the response function.
Explicit expressions for the first order correction terms for chaotic 
systems were developed by Gaspard and Alonso \cite{Alo93,Gas93} 
and by Vattay and Rosenqvist \cite{Vat94,Vat96,Ros94},
following two different approaches.
Vattay and Rosenqvist compute the corrections by solving the local
Schr\"odinger equation in the neighborhood of periodic orbits. 
They introduce a quantum generalization of the Gutzwiller formula
which contains these local eigenvalues.
The results of Refs.~\cite{Vat94,Vat96,Ros94} cannot directly be applied
to integrable systems, as the derivations are valid only for isolated 
periodic orbits.
To our knowledge, a general theory for $\hbar$ corrections to the 
Berry-Tabor formula does not yet exist.

Nevertheless, for the circle billiard we have succeeded in obtaining an 
explicit expression for the first order $\hbar$ corrections to the 
Berry-Tabor formula.
The calculations are quite lengthy and are therefore deferred to
Appendix \ref{appendix}.
Our final result for the first order $\hbar$ amplitude of the circle 
billiard in (\ref{g_hbar_series}) reads:
\begin{equation}
 {\cal A}_{\rm po}^{(1)}=
 {\cal A}_{\rm po}^{(0)}\ 
 {i\over 2}M_r
 \left({1\over 3 \sin\gamma}-{5\over 6\sin^3\gamma}\right) \; ,
\label{a1a0}
\end{equation}
with $\gamma\equiv\pi M_\varphi/M_r$ and ${\cal A}_{\rm po}^{(0)}$ 
the zeroth order amplitudes given by the Berry-Tabor formula. 
Using the zeroth order amplitudes from Eq.~(\ref{gw}), we finally obtain
\begin{equation}
\label{a1expl}
 {\cal A}_{\rm po}^{(1)} = \sqrt{w}
 \sqrt{\pi M_r} \, {2\sin^2\gamma-5\over 6\sin^{3/2}\gamma} \,
 e^{-i({3\over 2}M_r\pi-{\pi\over 4})} \; .
\end{equation}
As explained above our derivation of Eq.~(\ref{a1expl}) cannot be applied
to general integrable systems.
It will be an interesting task for the future to develop a {\em general} 
theory for the higher order $\hbar$ corrections to the Berry-Tabor formula.

\section{Harmonic inversion by filter-diagonalization}
\label{filter}
The quantization of the periodic orbit sum as well as the analysis
of quantum spectra in terms of the periodic orbits can be reformulated
as a harmonic inversion problem of formulae which have been introduced
in the previous Section \ref{POT}.
Before discussing these applications in Sections \ref{quantum} and \ref{poq} 
we will now briefly recapitulate the basic ideas and the technical tools of 
harmonic inversion by filter-diagonalization.
In Section \ref{hi1} we will start with the harmonic inversion of a single
function.
The equations will be generalized in Section \ref{hi2} to the harmonic 
inversion of cross-correlated signals.

\subsection{Harmonic inversion of a single function}
\label{hi1}
The harmonic inversion problem can be formulated as a nonlinear fit of a 
signal $C(t)$ to the form
\begin{equation}
\label{hiform}
 C(t) = \sum_kd_k e^{-i\omega_kt} \; ,
\end{equation}
where $d_k$ and $\omega_k$ are generally complex variational parameters.
Other than, e.g., in a simple Fourier transformation of the signal, there 
is no restriction to the closeness of the frequencies $\omega_k$. 
Solving $(\ref{hiform})$ will therefore yield a high resolution analysis of 
the signal $C(t)$. 
The signal length required for resolving the frequencies $\omega_k$ by 
harmonic inversion can be estimated to be 
\begin{equation}
\label{smax}
 t_{\rm max} \approx 4\pi\bar\rho(\omega), 
\end{equation}
where $\bar\rho(\omega)$ is the mean density of frequencies in the range of 
interest.

A method which has proven very useful for solving the harmonic inversion 
problem is the filter-diagonalization procedure \cite{Wal95,Man97}. 
This procedure allows us to compute the frequencies $\omega_k$ in any small 
interval $[\omega_{\rm min},\omega_{\rm max}]$ given. 
The idea is to consider the signal $C(t)$ on an equidistant grid 
\begin{equation}\label{dec_sig}
 c_n = C(n\tau) \; ; \quad n=0,1,2,\dots
\end{equation}
and to associate $c_n$ with an autocorrelation function of a suitable 
fictitious dynamical system, described by a complex symmetric effective 
Hamiltonian $H_{\rm eff}$:
\begin{equation}
\label{cn=}
 c_n = \left(\Phi_0|e^{-in\tau H_{\rm eff}}\Phi_0\right) \; .
\end{equation}
Here, the brackets denote a complex symmetric inner product $(a|b)=(b|a)$,
i.e., no complex conjugation of either $a$ or $b$.
The harmonic inversion problem can then be reformulated as solving the 
eigenvalue problem for the effective Hamiltonian $H_{\rm eff}$. 
The frequencies $\omega_k$ are the eigenvalues of the Hamiltonian
\begin{equation}
 H_{\rm eff}|\Upsilon_k) = \omega_k |\Upsilon_k) \; ,
\end{equation}
and the amplitudes are obtained from the eigenvectors $\Upsilon_k$:
\begin{equation}
 d_k = \left(\Phi_0|\Upsilon_k\right)^2 \; .
\end{equation}
The filter-diagonalization method solves this eigenvalue problem in a small 
set of basis vectors $\Psi_j$. 
The Hamiltonian and the initial state $\Phi_0$ do not have to be known 
explicitly but are given implicitly by the quantities $c_n$.
In detail, the procedure works as follows:

A small set of values $\varphi_j$ in the frequency interval of interest 
is chosen. 
The set must be larger than the number of frequencies in this interval. 
The values $\varphi_j$ are used to construct the small Fourier-type basis
\begin{equation}
 \Psi_j = \sum_{n=0}^M e^{in(\varphi_j-\tau H_{\rm eff})}\Phi_0 \; .
\end{equation}
The matrix elements of the evolution operator at a given time $p\tau$ in 
this basis can be expressed in terms of the quantities $c_n$:
\begin{equation}
   U^{(p)}_{jj'} \equiv \left(\Psi_j|e^{-ip\tau H_{\rm eff}} \Psi_{j'}\right)
 = \sum_{n=0}^M\sum_{n'=0}^Me^{i(n\varphi_j+n'\varphi_{j'})}c_{n+n'+p} \; .
\end{equation}
The frequencies $\omega_k$ are then obtained by solving the generalized 
eigenvalue problem
\begin{equation}
 {\bf U}^{(p)}{\bf B}_k = e^{-ip\tau\omega _k}{\bf B}_k \; .
\end{equation}
The amplitudes $d_k$ can be calculated from the eigenvectors and are given by
\begin{equation}
 d_k = \left(\sum_jB_{jk}\sum_{n=0}^Mc_ne^{in\varphi_j}\right)^2 \; .
\end{equation}
The values of $\omega_k$ and $d_k$ obtained by the above procedure should be 
independent of $p$. 
This condition can be used to identify non-converged frequencies by 
comparing the results for different values of $p$. 
The difference between the frequency values obtained for different $p$ 
can be used as a simple error estimate.

\subsection{Harmonic inversion of cross-correlated signals}
\label{hi2}
An important extension of the filter-diagonalization method for harmonic
inversion is the generalization to cross-cor\-re\-la\-tion functions 
\cite{Wal95,Nar97,Man98,Mai99b}. 
This extended method allows us to significantly reduce the signal length 
required to resolve the frequencies contained in the signal.
The idea is not to consider a single signal $C(t)$ as given 
in Eq.~(\ref{hiform}) but a set of cross-correlated signals
\begin{equation}\label{cross_sig}
 C_{\alpha\alpha'}(t) = \sum_k d_{\alpha\alpha',k}e^{-i\omega_kt} \; ; \quad
  \alpha,\alpha'=1,\dots,N
\end{equation}
with the restriction
\begin{equation}
\label{restrict}
 d_{\alpha\alpha',k} = b_{\alpha,k}b_{\alpha',k} \; .
\end{equation}
This set of signals considered on an equidistant grid
\begin{equation}
 c_{n\alpha\alpha'} = C_{\alpha\alpha'}(n\tau) \; ; \quad  n=0,1,2,\dots
\end{equation}
is now associated with a time cross-correlation function between an initial 
state $\Phi_\alpha$ and a final state $\Phi_{\alpha'}$:
\begin{equation}
\label{cnalpha=}
   c_{n\alpha\alpha'}
 = \left(\Phi_{\alpha'}|e^{-in\tau H_{\rm eff}}\Phi_\alpha\right) \; .
\end{equation}
Again, the frequencies $\omega_k$ are obtained as the eigenvalues of the 
effective Hamiltonian $H_{\rm eff}$. 
The amplitudes are now given by the relation
\begin{equation}
 b_{\alpha,k} = (\Phi_\alpha|\Upsilon_k) \; .
\end{equation}
In analogy to the procedure described in Section \ref{hi1}, this eigenvalue 
problem is solved in a small set of basis vectors $\Psi_{j\alpha}$ 
in order to obtain the frequencies in a given interval 
$[\omega_{\rm min},\omega_{\rm max}]$.

The advantage of the above procedure becomes evident if one considers the 
information content of the set of signals. Due to the restriction 
(\ref{restrict}), the $N\times N$ set of signals $C_{\alpha\alpha'}(t)$ 
may contain $N$ independent signals, which are known to possess the same 
frequencies $\omega_k$. This means that, at constant signal length, the 
matrix may contain $N$ times as much information about the frequencies as 
a single signal, provided that the whole set is inverted simultaneously. 
On the other hand, the information content is proportional to the signal 
length. This means that the signal length required to resolve the frequencies 
in a given interval is reduced by a factor of $N$.
This statement clearly holds only approximately and for small matrix 
dimensions $N$. However, a significant reduction of the required signal 
length can be achieved.

\section{High resolution analysis of quantum spectra}
\label{quantum}
Harmonic inversion is a powerful tool to calculate the classical periodic 
orbit contributions to the density of states from the quantum mechanical 
eigenvalues or from experimental spectra, thus delivering a high resolution 
analysis of the spectra in terms of the classical orbits.
The method allows us, e.g., to resolve clusters of orbits or to discover 
hidden ghost orbit contributions in the spectra, which would not be resolved 
by conventional Fourier analysis of the spectra \cite{Mai99a,Mai97a}.
Here, we will analyze the quantum spectra of the circle billiard
as a representative of integrable systems.
The analysis will verify the validity of the Berry-Tabor formula and its 
extensions to semiclassical matrix elements and higher order $\hbar$ 
corrections discussed in Section \ref{POT}.

\subsection{Leading order periodic orbit contributions to the trace formula}
\label{quantum1}
\subsubsection{General procedure}
\label{genproc2}
In this section we develop the general procedure for the analysis
of quantum spectra in terms of periodic orbits by harmonic inversion. 
This procedure is universal in the sense that it can be applied to 
both regular and chaotic systems.
We will apply it to the circle billiard as a representative of regular 
systems in the next section.

We start from the semiclassical density of states given by the Berry-Tabor
or the Gutzwiller formula.
As in Section \ref{POT}, we consider scaling systems where the density 
of states 
depends on the scaling parameter $w$ [$w=kR$ for the circle billiard],
i.e., $\rho(w)=-(1/\pi)\, {\rm Im}\ g(w)$ with $g(w)$ the semiclassical 
response function.
Both the Berry-Tabor and the  Gutzwiller formula give the oscillating part 
of the response function in the form
\begin{equation}\label{gsemiw}
 g^{\rm osc}(w) = \sum_{\rm po} {\mathcal A}_{\rm po} e^{{i}ws_{\rm po}} \; ,
\end{equation}
where the sum runs over all rational tori (regular systems) or all periodic 
orbits (chaotic systems) of the underlying classical system, respectively. 
Here, $S_{\rm po}$ is the action of the periodic orbit.
The form of the amplitude ${\mathcal A}_{\rm po}$ depends on whether the 
system is chaotic or regular and also contains phase information.

The exact quantum mechanical density of states is given by
\begin{equation}
\label{rhoqm}
 \rho_{\rm qm}(w) = \sum_k m_k \delta(w-w_k) \; ,
\end{equation}
where the $w_k$ are the exact quantum eigenvalues of the scaling parameter
and the $m_k$ are their multiplicities.
The analysis of the quantum spectrum in terms of periodic orbit 
contributions can now be reformulated as adjusting the exact quantum 
mechanical density of states (\ref{rhoqm}) to the semiclassical form
\begin{eqnarray}
\label{rhosemi2}
 \rho^{\rm osc}(w) &=& -{1\over\pi} \, {\rm Im}\ g^{\rm osc}(w) \nonumber \\
 &=& -{1\over 2\pi i}\ \sum_{\rm po}
     \left({\mathcal A_{\rm po}}e^{iws_{\rm po}}-
           {\mathcal A^\ast_{\rm po}}e^{-iws_{\rm po}}\right) \; .
\end{eqnarray}
If the amplitudes ${\mathcal A_{\rm po}}$ do not depend on $w$, the 
semiclassical density of states is of the form (\ref{hiform}) 
[here, with $w$ playing the role of $t$ and $s_{\rm po}$ playing the 
role of $\omega_k$].
That means, we have reformulated the problem of extracting the periodic 
contributions from the quantum spectrum as a harmonic inversion problem. 
In the fitting procedure, we ignore the non-oscillating, smooth
part of the density of states. This part does not fulfill the ansatz
(\ref{hiform}) of the harmonic inversion method and therefore
acts as a kind of noise, which will be separated from the oscillating
part of the ``signal'' by the harmonic inversion procedure.

In practice, in order to regularize the $\delta$ functions in (\ref{rhoqm}), 
we convolute both expressions (\ref{rhoqm}) and (\ref{rhosemi2}) with a 
Gaussian function,
\begin{equation}
 C_\sigma(w) = {1\over \sqrt{2\pi}\sigma}\int_{-\infty}^\infty
 \rho(w')e^{-(w-w')^2/2\sigma^2}dw' \; .
\end{equation}
In our calculations, we usually took the convolution width to be about 
three times the step width $\tau$ in the signal (\ref{dec_sig}).
Typical values are $\tau=\Delta w=0.002$ and $\sigma=0.006$.
The resulting quantum mechanical signal is
\begin{equation}
 C_{{\rm qm},\sigma}(w)={1\over \sqrt{2\pi}\sigma}
 \sum_k m_ke^{-(w-w_k)^2/2\sigma^2} \; ,
\end{equation}
and the corresponding semiclassical quantity reads
\begin{equation}
\label{cwsigma}
 C_{\sigma}(w) = -{1\over 2\pi i}\ \sum_{\rm po}
 \left({\mathcal A_{\rm po}}e^{iws_{\rm po}}-
 {\mathcal A^*_{\rm po}}e^{-iws_{\rm po}}\right)
 e^{-{1\over 2}\sigma^2 s_{\rm po}^2} \; .
\end{equation}

The above procedure still works if the amplitudes in (\ref{gsemiw}) are not 
independent of $w$ but possess a dependency of the form
\begin{equation}
 A_{\rm po} = w^\alpha a_{\rm po},
\end{equation}
which is, for example, the case for regular billiards. 
This dependency can be eliminated \cite{Mai98c} by replacing the 
semiclassical response function $g(w)$ with the quantity
\begin{equation}
\label{gstrich2}
 g'(w) = w^{-\alpha}g(w) 
       = w^{-\alpha}g_0(w)+\sum_{\rm po}{a_{\rm po}}e^{iws_{\rm po}} \; .
\end{equation}
When introducing the corresponding quantum mechanical response function
\begin{equation}
 g'_{\rm qm}(w) = \sum_k {m_k w_k^{-\alpha}\over w-w_k+i0}
\end{equation}
the procedure can be carried out for $\rho'(w)=(-1/\pi) {\rm Im}g'(w) $ 
as described above.

In addition to considering the pure density of states, the relations of 
Section \ref{matrixel} can be used to calculate the averages of classical 
quantities over the periodic orbits from the quantum diagonal matrix elements 
of the corresponding operators. 
If we start from the extended quantum response function (\ref{gaqme}),
including diagonal matrix elements of some operator $\hat A$,
the analysis of the signal should again yield the actions $s_{\rm po}$ 
as frequencies but with the amplitudes weighted with the classical
averages $\bar A_p$ of the corresponding classical quantities.
In the same way, we can also use the extended response function 
(\ref{gabqm}), which
includes diagonal matrix elements of two different operators.

\subsubsection{Application to the circle billiard}
\label{circle2}
For the circle billiard, the oscillating part $g^{\rm osc}(w)$ of the
semiclassical response function is given by Eq.~(\ref{gw}).
If one eliminates the dependency of the amplitudes on $w$ by defining 
\begin{equation}
\label{rhostrich}
 \rho'(w) = {1\over \sqrt w} \rho(w) \; ,
\end{equation}
the resulting expression for the density of states
\begin{eqnarray}
 \rho'(w)
 ={{1\over \sqrt{8\pi}}}
 \sum_{\bf M}m_{\bf M}{s_{\bf M}^{3/2}\over M_r^2}
\Bigl( & e^{i({3\over 2}M_r \pi-{\pi\over 4})} e^{-iws_{\bf M}}&
\nonumber \\
 + &  e^{-i({3\over 2}M_r \pi-{\pi\over 4})}e^{iws_{\bf M}} &\Bigr)
\label{rhostrichexpl}
\end{eqnarray}
is of the form (\ref{hiform}), here with $S_{\bf M}$ playing the role 
of $w_k$. 
The quantum mechanical quantity corresponding to (\ref{rhostrich}) is
\begin{equation}
\label{rhostrichqm}
 \rho'_{\rm qm}(w) = \sum_k{m_k\over \sqrt w_k}\delta(w-w_k) \; .
\end{equation}

In addition to analyzing the pure quantum spectrum of the circle
billiard, we also considered spectra weighted with diagonal matrix
elements of different operators (cf. Section \ref{matrixel}).
We used three different operators, viz.\
\begin{itemize}
\item the absolute value of the angular momentum $L$ as an example of 
a constant of motion,
\item the distance $r$ from the center as an example of a quantity 
which is no constant of motion,
\item the variance of the radius $\langle r^2\rangle-\langle r\rangle^2$ 
as an example using the relation (\ref{gab}) for products of operators.
\end{itemize}
The classical angular momentum $L$ is proportional to $w$, which means 
that when constructing the signal for $L$, $g(w)$ now has to be multiplied 
by $w^{-3/2}$ instead of $w^{-1/2}$ (cf. Eq.~(\ref{gstrich2})).

We calculated the scaled actions and classical amplitudes of the periodic 
orbits in the interval $s_{\bf M}\in [15,23]$.
The signal was constructed from the exact zeros of the Bessel functions, 
up to a value of $w_{\rm max}=500$.
The accuracy of results is improved if we cut off the lower part of the 
signal, using only zeros larger than $w_{\rm min}=300$.
A possible explanation for this is that the low zeros are in a sense 
``too much quantum'' for the semiclassical periodic orbit sum.

\def\baselinestretch{1}
\begin{figure}
\vspace{12cm}       
\includegraphics{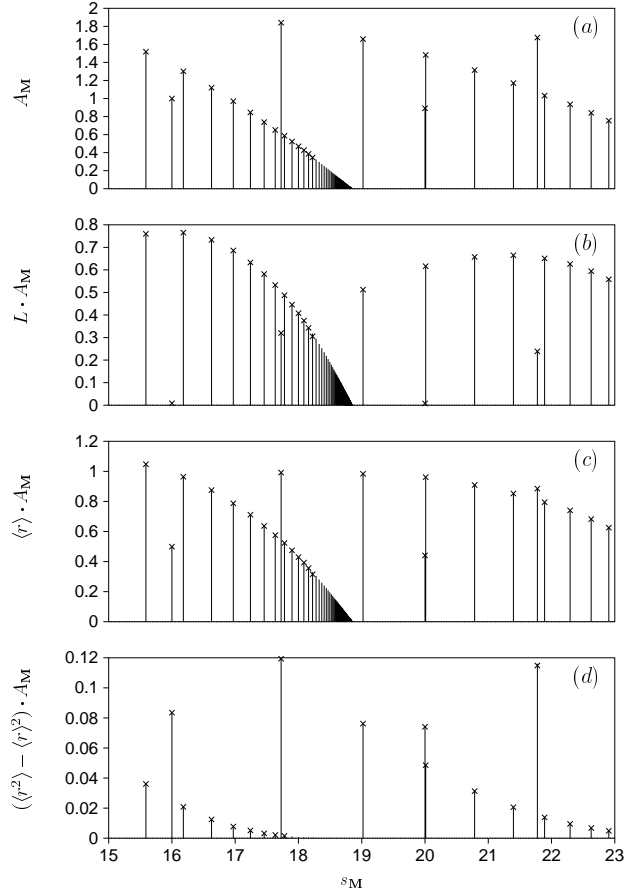}
\caption{Periodic orbit contributions to the trace formula calculated from 
the quantum spectrum of the circle billiard, including different operators. 
The quantities plotted are the classical amplitudes 
$A_{\bf M}=m_{\bf M}{s_{\bf M}^{3/2}/ M_r^2}$ times the classical 
averages of the 
operators indicated, versus the scaled actions of the orbits. Crosses: 
values obtained from the quantum spectrum by harmonic inversion. 
Solid lines: values obtained from classical calculations.
$L$: angular momentum in units of $\hbar w$ 
($w$: scaling parameter), $r$: distance from the
center in units of the radius $R$ of the billiard.}
\label{fig3}
\end{figure}
\def\baselinestretch{2}
Figure \ref{fig3} shows the results of our calculation. The positions 
of the solid lines are the scaled actions of the classical periodic orbits, 
their heights are the classical amplitudes 
$m_{\bf M}{s_{\bf M}^{3/2}/M_r^2}$ times the respective averaged 
classical quantity. 
The crosses are the results obtained by harmonic inversion of the signal 
constructed from the zeros of the Bessel functions. There is an excellent
agreement between the spectra, illustrating the validity of the Berry-Tabor
formula and its extension to semiclassical matrix elements discussed in
Section \ref{matrixel}.
The examined interval contains an accumulation point of orbits ($s=6\pi$).
Here, only those orbits were resolved which were still sufficiently isolated. 

It might be surprising that, although the Berry-Tabor formula only gives
a semiclassical approximation to the density of states and we started
from the exact quantum mechanical density, our results for the periodic 
orbit contributions are exact and do not show any deviations due to the 
error of the semiclassical approximation. 
The reason for this will become obvious in the following Section.

\subsection{Higher order $\hbar$ corrections to the trace formula}
\label{hbar2}
An interesting application of the method described in the previous
Section \ref{quantum1} is the determination of higher order $\hbar$ 
contributions to the periodic orbit sum. 
The higher orders can be obtained by analysis of the difference spectrum 
between the exact quantum and semiclassical eigenvalues, as we will show 
below.

As explained in Section \ref{hbarsec1}, the Berry-Tabor formula for 
integrable systems as well as the Gutzwiller formula for chaotic systems 
are the leading order terms of an expansion of the density of states in 
terms of $\hbar$. For scaling systems, this expansion can be put in
the form (cf.\ (\ref{g_hbar_series}))
\begin{equation}
\label{g_hbar_series3}
   g^{\rm osc}(w)
 = \sum_{n=0}^\infty g_n(w)
 = \sum_{n=0}^\infty {1\over w^{n}} \sum_{\rm po} 
   {\cal A}_{\rm po}^{(n)} e^{is_{\rm po}w} \; .
\end{equation}
Provided that the amplitudes ${\cal A}_{\rm po}^{(n)}$ in 
(\ref{g_hbar_series3}) do not depend on $w$, only the zeroth order term 
fulfills the ansatz (\ref{hiform}) for the harmonic inversion procedure 
with constant amplitudes and frequencies. In systems like 
regular billiards, where the amplitudes possess a $w$ dependence of the 
form ${\cal A}_{\rm po}^{(n)}=w^\alpha {a}_{\rm po}^{(n)}$, the same 
argumentation holds if we consider $g'(w)=w^{-\alpha}g(w)$ instead of 
$g(w)$ (cf. Section \ref{genproc2}).
This is the reason why the analysis of the quantum spectrum 
yields exactly the lowest order amplitudes ${\cal A}_{\rm po}^{(0)}$, 
without any deviations due to the semiclassical error: As the higher 
order terms do not fulfill the ansatz, the ${\cal A}_{\rm po}^{(0)}$ 
are the best fit for the amplitudes. The higher oder terms have similar 
properties as a weak noise and are separated from the ``true'' signal 
by the harmonic inversion procedure.

Although the direct analysis of the quantum spectrum only yields the 
lowest order amplitudes, higher order corrections can still be 
extracted from the spectrum by harmonic inversion. Assume that the 
exact eigenvalues $w_k$ and their $(n-1)^{\rm st}$ order approximations 
$w_{k,n-1}$ are given. 
We can then calculate the difference between the exact quantum mechanical
and the $(n-1)^{\rm st}$ order response function
\begin{equation}
\label{Dg_n}
   g^{\rm qm}(w) - \sum_{j=0}^{n-1} g_j(w)
 = \sum_{j=n}^\infty g_j(w)
 = \sum_{j=n}^\infty {1\over w^{j}} \sum_{\rm po} 
   {\cal A}_{\rm po}^{(j)} e^{is_{\rm po}w} .
\end{equation}
The leading order terms in (\ref{Dg_n}) are $\sim w^{-n}$, i.e.,
multiplication with $w^n$ yields
\begin{equation}
\label{G_n}
   w^n\left[g^{\rm qm}(w) - \sum_{j=0}^{n-1} g_j(w)\right]
 = \sum_{\rm po} {\cal A}_{\rm po}^{(n)} e^{is_{\rm po}w}
   + {\cal O}\left({1\over w}\right) .
\end{equation}
In (\ref{G_n}) we have restored the functional form (\ref{hiform}).
The harmonic inversion of the function (\ref{G_n}) will now provide the 
periods $s_{\rm po}$ and the $n^{\rm th}$ order amplitudes 
${\cal A}_{\rm po}^{(n)}$ of the $\hbar$ expansion (\ref{g_hbar_series3}).

In practice, we will follow the procedure outlined in Section \ref{genproc2}
to construct a smooth signal, i.e., we consider the densities of states 
$\rho(w)=-(1/\pi)\, {\rm Im}\, g(w)$ rather than the response functions 
$g(w)$, and smoothen the signal by convoluting it with a Gaussian function.

For the circle billiard, the exact quantum eigenvalues are given by the 
condition (\ref{jl=0}), while the zeroth order eigenvalues are 
equal to the EBK eigenvalues given by (\ref{EBKev}) 
(cf. Section \ref{Berry-Tabor}). From the difference between the exact 
and the semiclassical density of states, we can calculate the amplitudes 
${\cal A}_{\rm po}^{(1)}$ of the first order correction to the trace formula. 

\def\baselinestretch{1}
\begin{figure}
\vspace{5.5cm}       
\includegraphics{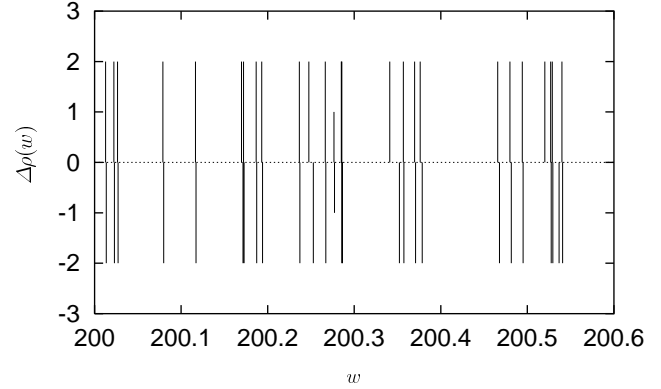}
\caption{Part of the difference spectrum 
$\Delta\rho_{\rm qm}(w)-\Delta\rho_{\rm EBK}(w)$
between the exact quantum mechanical and the semiclassical density of states. 
The absolute values of the peak heights mark the multiplicities 
of the states.}
\label{fig4}
\end{figure}
\def\baselinestretch{2}
We analyzed the difference spectrum between exact and EBK eigenvalues of 
the circle billiard in the range $100<w<500$. Figure \ref{fig4} shows
a small part of this difference spectrum.
The results of the harmonic inversion of the spectrum are presented in 
Figure \ref{fig5}. 
\def\baselinestretch{1}
\begin{figure}
\vspace{5.5cm}       
\includegraphics{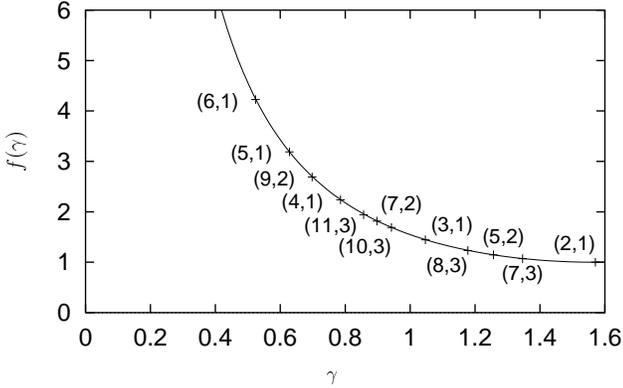}
\caption{First order $\hbar$ corrections to the Berry-Tabor formula for the 
circle billiard. Crosses: amplitudes 
$(2/\sqrt{\pi M_r})\, |{\cal A}_{\rm po}^{(1)}|$
 obtained by harmonic inversion. The orbits are labeled by the numbers 
$(M_r,M_\varphi)$. Solid line: theoretical curve
$f(\gamma)=(5-2\sin^2\gamma)/ (3\sin^{3/2}\gamma)$.}
\label{fig5}
\end{figure}
\def\baselinestretch{2}
The crosses mark the values
\begin{equation}
 f(\gamma) \equiv {2\over\sqrt{\pi M_r}}\,
 {1\over \sqrt{w}}\, |{\cal A}_{\rm po}^{(1)}| \; ,
\end{equation}
with $\gamma=\pi{M_\varphi/M_r}$ which we 
obtained for the periodic orbits by harmonic inversion of the difference 
spectrum. The crosses are labeled with the numbers $(M_r,M_{\varphi})$ of 
the orbits.
The solid line in Fig.~\ref{fig5} is the theoretical curve
\begin{equation}
 f(\gamma) = {5-2\sin^2\gamma\over 3\sin^{3/2}\gamma} \; ,
\end{equation}
which results from our analytical expression (\ref{a1expl}) for the first
order amplitudes discussed in Section \ref{hbarsec1}. 
The results obtained by harmonic inversion are in excellent agreement 
with the theoretical curve, which clearly illustrates the validity of
Eq.~(\ref{a1expl}).

\section{Periodic orbit quantization}
\label{poq}
\subsection{General procedure}
\label{genproc}
We now turn to the problem of extracting eigenvalues from the periodic
orbit sum.
We will demonstrate that the harmonic inversion procedure, which has already
been successfully applied to extract the eigenvalues of chaotic systems
\cite{Mai97b,Mai98}, can be used for integrable systems as well when starting
from the Berry-Tabor formula.

As previously (see Section \ref{POT}), we consider scaling systems and start 
from the response function 
\begin{equation}
\label{gsemiw2}
 g(w) = g_0(w) + \sum_{\rm po}{\mathcal A_{\rm po}}e^{iws_{\rm po}} \; ,
\end{equation}
depending on the scaling parameter $w$.
The amplitudes ${\mathcal A_{\rm po}}$ are those of the Berry-Tabor
or the Gutzwiller formula for regular and chaotic systems, respectively.
The periodic orbit sum in (\ref{gsemiw2}) usually does not converge, or,
at least, the convergence will be very slow.
In practice, especially for chaotic systems, only the orbits with small 
scaled actions will be known.
Nevertheless, the eigenvalues of the scaling parameter can still be 
extracted from the periodic orbit sum.
The central idea is to adjust Eq.~(\ref{gsemiw2}), with the sum including 
periodic orbits up to a finite action $s_{\rm max}$, to the functional form 
of the corresponding quantum mechanical response function
\begin{equation}
\label{gqmw}
 g_{\rm qm}(w) = \sum_k {m_k\over w-w_k+i0} \; .
\end{equation}
This fitting problem cannot be solved directly, but
can be reformulated as a harmonic inversion problem \cite{Mai97b,Mai98}.
The first step of the reformulation is a Fourier transformation of the 
response functions with respect to $w$:
\begin{equation}
 C(s) = {1\over 2\pi}\int_{-\infty}^\infty g(w)e^{-isw}dw \; .
\end{equation}
In the semiclassical response function, we only consider the oscillating 
part of $g(w)$.
The smooth part, which does not possess a suitable form for the harmonic 
inversion method, would only give a contribution to the signal for very 
small $s$.
Assuming that the amplitudes in (\ref{gsemiw2}) do not depend on $w$, the 
result of the Fourier transformation is
\begin{eqnarray}
 C(s)&=&
 \sum_{\rm po}{\mathcal A_{\rm po}}\delta(s-s_{\rm po}) \; ,
\label{cs1} \\
 C_{\rm qm}(s)&=&-i\sum_k m_k e^{-isw_k}.
\label{cqm1}
\end{eqnarray}
Like in Section \ref{genproc2} we convolute the signals (\ref{cs1}) and 
(\ref{cqm1}) with a Gaussian function with width $\sigma$, resulting in
\begin{eqnarray}
 C_\sigma(s)&=&
 {1\over \sqrt{2\pi}\sigma}\sum_{\rm po}{\mathcal A_{\rm po}}
 e^{-(s-s_{\rm po})^2/2\sigma^2} \; ,
\label{csigmas1}\\
 C_{{\rm qm},\sigma}(s)
 &=&-i\sum_k m_k
 e^{-{1\over 2}\sigma^2w_k^2}e^{-isw_k} \; .
\label{csigmaqm1}
\end{eqnarray}
Typical values of the convolution width are $\sigma=0.006$ for signals
with step width $\Delta s=0.002$.
The eigenvalues of the scaling parameter are now obtained by adjusting
the signal $C_\sigma(s)$ to (\ref{csigmaqm1}), which is of the functional 
form (\ref{hiform}).
The frequencies $w_k$ obtained by harmonic inversion of the signal 
(\ref{csigmas1}) are the eigenvalues of the scaling parameter $w$; 
from the amplitudes $d_k$, 
the multiplicities $m_k$ can be calculated.

Like the general procedure for analyzing quantum spectra 
(see Section \ref{genproc2}), the above procedure still works if the 
amplitudes in (\ref{gsemiw2}) are not independent of $w$ but possess 
a dependency of the form
\begin{equation}
 {\mathcal A}_{\rm po} = w^\alpha a_{\rm po} \; .
\end{equation}
We can again eliminate this dependency by replacing $g(w)$ with the quantity
\begin{equation}
 g'(w) = w^{-\alpha}g(w) \; .
\end{equation}
The semiclassical signal is then given by
\begin{equation}
\label{ca}
 C_\sigma(s)= {1\over \sqrt{2\pi}\sigma}\sum_{\rm po}a_{\rm po}
 e^{-(s-s_{\rm po})^2/2\sigma^2} \; ,
\end{equation}
and the corresponding quantum mechanical signal reads
\begin{equation}
\label{caqm}
 C_{{\rm qm},\sigma}(s) = -i\sum_k m_k w^{-\alpha}_k
 e^{-{1\over 2}\sigma^2w_k^2}e^{-isw_k} \; .
\end{equation}

\subsection{Semiclassical eigenvalues of the circle billiard}
\label{circle3}
\subsubsection{Construction of the periodic orbit signal}
\label{constrsig}
The semiclassical response function of the circle billiard is given by 
Eq.~(\ref{gw}). 
The amplitudes in (\ref{gw}) are proportional to $w^{1/2}$.
As described in Section \ref{genproc}, we eliminate this dependency 
on $w$ by introducing the quantity
\begin{equation}
\label{gstrich}
 g'(w) = w^{-1/2} g(w) \; .
\end{equation}
Applying Eqs.~(\ref{ca}) and (\ref{caqm}), we now obtain the semiclassical 
and the corresponding exact quantum signal for the circle billiard:
\begin{eqnarray}
 C_\sigma(s) &=& {e^{-i{\pi\over 4}}\over 2\sigma} \sum_{\bf M}
 m_{\bf M}{s_{\bf M}^{3/2}\over M_r^2}
 e^{-i{3\over 2}M_r\pi} e^{-(s-s_{\bf M})^2/2\sigma^2} \; ,\nonumber\\
\label{csigmas} \\
 C_{{\rm qm},\sigma}(s) &=& -i\sum_k {m_k\over \sqrt w_k}
 e^{-{1\over 2}\sigma^2w_k^2}e^{-isw_k} \; .
\label{csigmaqm}
\end{eqnarray}
Eq.~(\ref{csigmaqm}) possesses the functional form (\ref{hiform}) with 
\begin{equation}
\label{dk=}
 d_k = -i{m_k\over \sqrt w_k} e^{-{1\over 2}\sigma^2w_k^2} \; .
\end{equation}
Applying the harmonic inversion method to the signal (\ref{csigmas}) 
should yield the eigenvalues of $w$ as frequencies, with the amplitudes 
given by Eq.~(\ref{dk=}).

\subsubsection{Results for the lowest eigenvalues}
\label{results}
We calculated the eigenvalues of the scaling parameter $w=kR$ for the lowest 
states of the circle billiard from a signal of length $s_{\rm max}=150$.
For the construction of the signal, we chose a minimum length for the 
sides of the periodic orbits as cut-off criterion at the accumulation 
points (cf. Fig.~\ref{fig2}). 
We observed that the results were nearly independent of the 
choice of this parameter, as long as the minimum length was not chosen 
too large.

\def\baselinestretch{1}
\begin{table}
\caption{Lowest eigenvalues $w_k$ and multiplicities $m_k$ 
of the scaling 
parameter $w=kR$ of the circle billiard. $w_{\rm ex}$ and $m_{\rm ex}$: 
exact quantum values. $w_{\rm EBK}$: EBK eigenvalues.
$w_{\rm hi}$ and $m_{\rm hi}$: values obtained by harmonic inversion of a 
signal of length $s_{\rm max}=150$. The numbers $n$ and $m$ are the radial
 and angular momentum quantum numbers. The nearly degenerate eigenvalues at 
$\approx 11.0$ and $\approx 13.3$ were not resolved.}
\label{tab1}
\begin{center}
\begin{tabular}{rrrrrrr}
\hline\noalign{\smallskip}
$n$ & $m$ & $w_{\rm ex}$ & $w_{\rm EBK}$ & $m_{\rm ex}$ & $w_{\rm hi}$ & $m_{\rm hi}$\\
\noalign{\smallskip}\hline\noalign{\smallskip}
0 &  0 &   2.404826   &    2.356194   &   1   &    2.356204   &    1.0005   \\
0 &  1 &   3.831706   &    3.794440   &   2   &    3.794444   &    1.9983   \\ 
0 &  2 &   5.135622   &    5.100386   &   2   &    5.100391   &    1.9996   \\ 
1 &  0 &   5.520078   &    5.497787   &   1   &    5.497785   &    0.9988   \\ 
0 &  3 &   6.380162   &    6.345186   &   2   &    6.345191   &    2.0000   \\ 
1 &  1 &   7.015587   &    6.997002   &   2   &    6.997006   &    2.0001   \\ 
0 &  4 &   7.588342   &    7.553060   &   2   &    7.553065   &    1.9992   \\ 
1 &  2 &   8.417244   &    8.400144   &   2   &    8.400149   &    1.9998   \\ 
2 &  0 &   8.653728   &    8.639380   &   1   &    8.639404   &    0.9987   \\ 
0 &  5 &   8.771484   &    8.735670   &   2   &    8.735677   &    2.0013   \\ 
1 &  3 &   9.761023   &    9.744628   &   2   &    9.744632   &    1.9999   \\ 
0 &  6 &   9.936110   &    9.899671   &   2   &    9.899675   &    1.9999   \\ 
2 &  1 &  10.173468   &   10.160928   &   2   &   10.160932   &    2.0000   \\ 
1 &  4 &  11.064709   &   11.048664   &   2   &               &             \\   
0 &  7 &  11.086370   &   11.049268   &   2   &
\raisebox{1.5ex}[-1.5ex]{11.048968}   &    \raisebox{1.5ex}[-1.5ex]{4.0012} \\ 
2 &  2 &   11.619841   &   11.608251   &   2   &   11.608256   &    2.0006   \\  
3 &  0 &  11.791534   &   11.780972   &   1   &   11.780978   &    1.0001   \\
0 &  8 &  12.225092   &   12.187316   &   2   &   12.187319   &    1.9993   \\ 
1 &  5 &  12.338604   &   12.322723   &   2   &   12.322724   &    2.0000   \\  
2 &  3 &  13.015201   &   13.004166   &   2   &   13.004168   &    1.9997   \\ 
3 &  1 &  13.323692   &   13.314197   &   2   &               &             \\   
0 &  9 &  13.354300   &   13.315852   &   2   &
\raisebox{1.5ex}[-1.5ex]{13.315045}    &   \raisebox{1.5ex}[-1.5ex]{4.0287} \\ 
1 &  6 &  13.589290   &   13.573465   &   2   &   13.573465   &    2.0000   \\
2 &  4 &  14.372537   &   14.361846   &   2   &   14.361849   &    1.9994   \\
0 & 10 &  14.475501   &   14.436391   &   2   &   14.436395   &    2.0006   \\ 
3 &  2 &  14.795952   &   14.787105   &   2   &   14.787076   &    1.9909   \\ 
1 &  7 &  14.821269   &   14.805435   &   2   &   14.805453   &    2.0066   \\ 
4 &  0 &  14.930918   &   14.922565   &   1   &   14.922569   &    1.0001   \\ 
\noalign{\smallskip}\hline
\end{tabular}
\end{center}
\end{table}
\def\baselinestretch{2}
Table \ref{tab1} presents the semiclassical eigenvalues $w_{\rm hi}$ and
multiplicities $m_{\rm hi}$ obtained by harmonic inversion of the periodic
orbit signal (\ref{csigmas}).
For comparison, the exact quantum mechanical and the EBK results are also 
given in Table \ref{tab1}.
The eigenvalues obtained by harmonic inversion clearly reproduce the EBK 
eigenvalues within an accuracy of $10^{-4}$ or better. 
The deviation of the $w_{\rm hi}$ from the EBK eigenvalues is 
significantly smaller than the error of the semiclassical approximation.
The improvement of the semiclassical quantization by including higher order
$\hbar$ corrections to the periodic orbit sum will be discussed in Section
\ref{hbar1}.

Note that for calculating the eigenvalues of the circle billiard 
by a direct evaluation of the periodic orbit sum,
a huge number of periodic orbit terms 
is required, e.g., orbits with maximum length $s_{\rm max}=30\,000$ were 
included in Ref.~\cite{Rei96}.
We obtained similar results using only orbits up to length $s_{\rm max}=150$.
This demonstrates the high efficiency of the harmonic inversion method
in extracting eigenvalues from the periodic orbit sum.
The efficiency can even be further increased with the help of  
cross-correlated periodic orbit sums as will be demonstrated in 
Section \ref{crosscorr}.

In Table~\ref{tab1} the exact multiplicities $m_{\rm ex}$ of eigenvalues 
and the multiplicities $m_{\rm hi}$ obtained by harmonic inversion also
agree to very high precision.
The deviations are one or two orders of magnitude larger than those 
in the frequencies, which reflects the fact that, with the harmonic 
inversion method using filter-diagonalization, the amplitudes usually 
converge more slowly than the frequencies.

In the frequency interval shown, there are two cases of nearly degenerate 
frequencies which have not been resolved by harmonic inversion of the
periodic orbit signal with $s_{\rm max}=150$.
The harmonic inversion yielded only one frequency, which is the average 
of the two nearly degenerate ones, with the amplitudes of the two added. 
These nearly degenerate states can be resolved when the signal length is 
increased to about $s_{\rm max}=500$ or with the help of cross-correlated 
periodic orbit sums (see Section \ref{crosscorr}).

\subsubsection{Semiclassical matrix elements}
\label{matrixel2}
Using the extended periodic orbit sums discussed in Section \ref{matrixel}, 
we can now also calculate semiclassical diagonal matrix elements for the
circle billiard.

Following the procedure described in Section \ref{genproc}, a semiclassical 
signal can be constructed from the extended response function, the analysis
of which should again yield the eigenvalues $w_k$ as frequencies but with 
the amplitudes weighted with the diagonal matrix elements.
As examples, we used the same operators as in Section \ref{circle2} to
calculate the diagonal matrix elements $\langle L\rangle$, $\langle r\rangle$,
and the variance of the radius, $\langle r^2\rangle-\langle r\rangle^2$.

\def\baselinestretch{1}
\begin{figure}
\vspace{12.2cm}       
\includegraphics{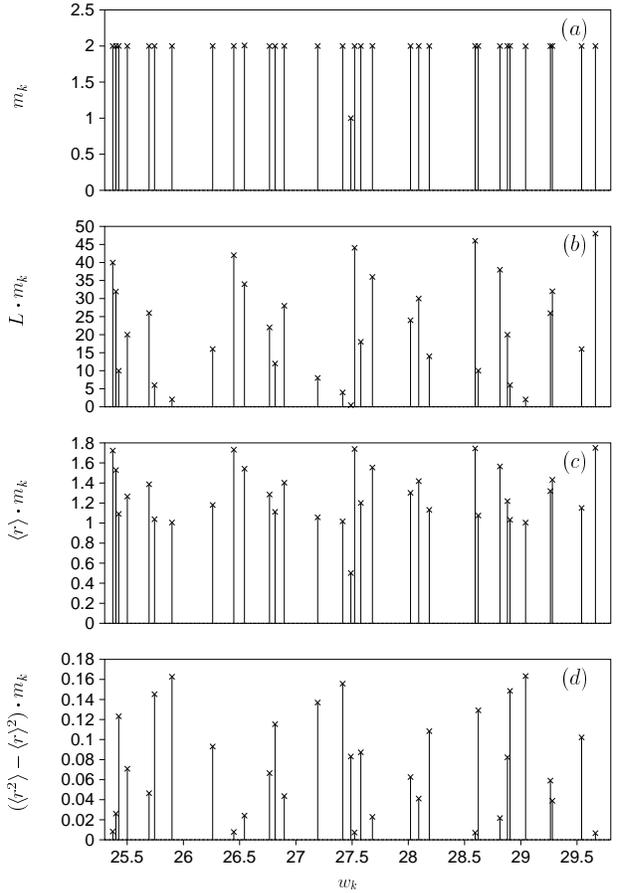}
\caption{Circle billiard: Results of the harmonic inversion of the 
semiclassical signal constructed from the periodic orbits sum including 
different operators. The quantities plotted are the diagonal matrix 
elements of the operators indicated times the multiplicities $m_k$, versus the 
eigenvalues $w_k$ of the scaling parameter. Crosses: results of the 
harmonic inversion procedure. Solid lines: semiclassical matrix elements 
obtained from EBK theory and EBK eigenvalues.
$L$: angular momentum in units of $\hbar$, $r$: distance from the
center in units of the radius $R$ of the billiard.}
\label{fig6}
\end{figure}
\def\baselinestretch{2}
Figure \ref{fig6} shows the results in the range $25\le w \le 30$.
For comparison, Fig.~\ref{fig6}a presents the spectrum for the identity
operator.
The positions of the solid lines are the EBK eigenvalues, their heights 
are the semiclassical matrix elements obtained from EBK theory
times the multiplicities.
The crosses are the results of the harmonic inversion of a signal of 
length $s_{\rm max}=300$.
The diagrams show excellent agreement between the results obtained by 
harmonic inversion and EBK torus quantization.
This is even the case for the variance of $r$, which is a very small quantity.

For the states shown in Fig.~\ref{fig6} we have also compared the 
semiclassical to the exact quantum mechanical matrix elements.
The agreement is also excellent.
The deviations between the semiclassical and quantum matrix elements are
typically of the order of $\sim 10^{-3}$, which can well be understood by the 
semiclassical approximation.

\subsection{Higher order $\hbar$ corrections}
\label{hbar1}
The eigenvalues of the circle billiard obtained in the previous Section
\ref{circle3} are not the exact quantum mechanical eigenvalues but 
semiclassical approximations for the reason that the Berry-Tabor and the 
Gutzwiller formula are only the leading order terms of an expansion of 
the density of states in terms of $\hbar$ (see Section \ref{hbarsec1}).
We will now demonstrate how to obtain corrections to the semiclassical 
eigenvalues from the $\hbar$ expansion (\ref{g_hbar_series}) of the 
periodic orbit sum
\[
 g^{\rm osc}(w) = \sum_{n=0}^\infty g_n(w)
 = \sum_{n=0}^\infty {1\over w^{n}} \sum_{\rm po} 
 {\cal A}_{\rm po}^{(n)} e^{is_{\rm po}w} \; ,
\]
with $w=\hbar_{\rm eff}^{-1}$ an effective inverse Planck constant
(see Eq.~(\ref{hbar_eff})).
The amplitudes ${\cal A}_{\rm po}^{(0)}$ are those of the Berry-Tabor or 
Gutzwiller formula.
For $n>0$, the amplitudes ${\cal A}_{\rm po}^{(n)}$ (including also phase 
information) give the $n^{\rm th}$ order corrections $g_n(w)$ to the 
response function $g^{\rm osc}(w)$.
For simplicity, we will assume in the following that the amplitudes 
${\cal A}_{\rm po}^{(n)}$ in (\ref{g_hbar_series}) do not depend on $w$. 
Again, in systems where the amplitudes possess a $w$ dependence of the 
form ${\cal A}_{\rm po}^{(n)}=w^\alpha {a}_{\rm po}^{(n)}$, the same line
of arguments holds if we consider $g'(w)=w^{-\alpha}g(w)$ instead of 
$g(w)$ (cf. Section \ref{genproc}).

For periodic orbit quantization the zeroth order contributions 
${\cal A}_{\rm po}^{(0)}$ are usually considered only.
As demonstrated in Section \ref{genproc} (see Eqs.~(\ref{cs1}) and 
(\ref{cqm1})), the Fourier transform of the principal periodic orbit sum
\[
 C_0(s) = \sum_{\rm po}{\cal A}_{\rm po}^{(0)} \delta(s-s_{\rm po})
\]
is adjusted by application of the harmonic inversion technique to the 
functional form of the exact quantum expression
\[
 C_{\rm qm}(s) = -i \sum_k m_k e^{-iw_ks} \; ,
\]
with $\{w_k,m_k\}$ the eigenvalues and multiplicities.

For $n\ge 1$, the asymptotic expansion (\ref{g_hbar_series}) of the 
semiclassical response function suffers from the singularities 
at $w=0$.
It is therefore not appropriate to harmonically invert the Fourier transform 
of (\ref{g_hbar_series}) as a whole, although the Fourier transform formally 
exists.
This means that the method of periodic orbit quantization by harmonic 
inversion cannot straightforwardly be extended to the $\hbar$ expansion of
the periodic orbit sum.
Instead, we will calculate the correction terms to the semiclassical 
eigenvalues separately, order by order \cite{Mai98c}.

Let us assume that the $(n-1)^{\rm st}$ order approximations $w_{k,n-1}$ 
to the semiclassical eigenvalues have already been obtained and the 
$w_{k,n}$ are to be calculated.
The difference between the two subsequent approximations to the quantum
mechanical response function reads
\begin{eqnarray}
     g_{n}(w)
 &=& \sum_k \left({m_k\over w-w_{k,n}+i0}
          - {m_k\over w-w_{k,n-1}+i0}\right)
     \nonumber \\
 &\approx& \sum_k{m_k\Delta w_{k,n}\over (w-\bar w_{k,n}+i0)^2} \; ,
\label{g_n}
\end{eqnarray}
with $\bar w_{k,n}={1\over 2}(w_{k,n}+w_{k,n-1})$ and 
$\Delta w_{k,n}=w_{k,n}-w_{k,n-1}$.
Integration of (\ref{g_n}) and multiplication by $w^n$ yields
\begin{equation}
 {\cal G}_{n}(w) = w^n \int g_{n}(w)dw
 = \sum_k {-m_kw^n\Delta w_{k,n}\over w-\bar w_{k,n}+i0} \; ,
\label{g_int_qm}
\end{equation}
which has the functional form of a quantum mechanical response function 
but with residues proportional to the $n^{\rm th}$ order corrections 
$\Delta w_{k,n}$ to the semiclassical eigenvalues.
The semiclassical approximation to (\ref{g_int_qm}) is obtained from 
the term $g_{n}(w)$ in the periodic orbit sum (\ref{g_hbar_series}) by 
integration and multiplication by $w^n$, i.e.,
\begin{eqnarray}
     {\cal G}_{n}(w)
 &=& w^n \int g_{n}(w)dw \nonumber \\
 &=& -i\sum_{\rm po} {1\over s_{\rm po}}
     {\cal A}_{\rm po}^{(n)} e^{iws_{\rm po}}
     + {\cal O}\left(1\over w\right) \; .
\label{g_int_sc}
\end{eqnarray}
We can now Fourier transform both (\ref{g_int_qm}) and (\ref{g_int_sc}),
and obtain ($n\ge 1$)
\begin{eqnarray}
     C_{n}(s)
 &\equiv& {1\over 2\pi}\int_{-\infty}^{+\infty}{\cal G}_{n}(w)e^{-iws}dw
     \nonumber  \\
\label{Cn_qm}
 &=& i\sum_k m_k (\bar w_k)^n\Delta w_{k,n} e^{-i\bar w_ks} \\
\label{Cn_sc}
 &\stackrel{\rm h.i.}{=}&
     -i\sum_{\rm po}{1\over s_{\rm po}}{\cal A}_{\rm po}^{(n)}
     \delta(s-s_{\rm po}) \; .
\end{eqnarray}
Equations (\ref{Cn_qm}) and (\ref{Cn_sc}) imply that the $\hbar$ expansion 
of the semiclassical eigenvalues can be obtained, order by order, by 
harmonic inversion (h.i.) of the periodic orbit signal in (\ref{Cn_sc}) to 
the functional form of (\ref{Cn_qm}).
[In practice, we will again convolute both expressions with a Gaussian 
function (cf.\ Section \ref{genproc}) in order to regularize the $\delta$ 
functions in (\ref{Cn_sc}).]
The frequencies $\bar w_k$ of the periodic orbit signal (\ref{Cn_sc}) are the 
semiclassical eigenvalues, averaged over different orders of $\hbar$.
Note that the accuracy of these values does not necessarily
increase with increasing order $n$.
We indicate this in (\ref{Cn_qm}) by omitting the index $n$ at the 
eigenvalues $\bar w_k$.
Our numerical calculations for the first order $\hbar$ corrections 
show that, in practice, the frequencies $\bar w_k$
we obtain are approximately equal to the zeroth order $\hbar$ eigenvalues
rather than the exact average between zeroth and first oder eigenvalues.
The corrections $\Delta w_{k,n}$ to the eigenvalues are not obtained
from the frequencies, but from 
the {amplitudes}, $m_k(\bar w_k)^n\Delta w_{k,n}$, of the periodic 
orbit signal. 

We will now apply the above technique to the circle billiard to obtain the 
first order corrections to the semiclassical eigenvalues obtained in Section 
\ref{results}. 
In Section \ref{circle}, we derived the zeroth order amplitudes of 
the circle billiard (cf.\ Eq.\ (\ref{gw})):
\begin{equation}
  {1\over\sqrt w}{\cal A}_{\rm po}^{(0)}
= \sqrt{{\pi\over 2}} m_{\bf M}{s_{\bf M}^{3/2}\over M_r^2}
  e^{-i({3\over 2}M_r\pi+{\pi\over 4})} \; ,
\end{equation}
with $s_{\bf M}$ and $m_{\bf M}$ the action and multiplicity of the orbit,
respectively.
The first order amplitudes are given by (cf.\ Sections \ref{hbarsec1}
and \ref{hbar2}):
\begin{equation}
\label{apo1}
 {1\over\sqrt w}{\cal A}_{\rm po}^{(1)}
 =\sqrt{\pi M_r} \, {2\sin^2\gamma-5\over 6\sin^{3/2}\gamma} \,
 e^{-i({3\over 2}M_r\pi-{\pi\over 4})} \; ,
\end{equation}
with $\gamma\equiv\pi M_\varphi/M_r$.

\def\baselinestretch{1}
\begin{figure}
\vspace{6.2cm}       
\includegraphics{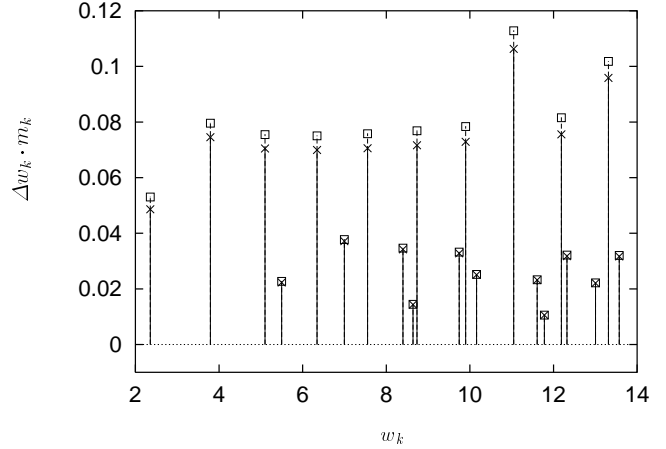}
\caption{Correction terms to the semiclassical eigenvalues of the circle 
billiard. Squares: corrections $\Delta w_{k,1}= w_{k,1}-w_{k,0}$ between 
first and zeroth order approximations (times multiplicities) obtained by 
harmonic inversion. Crosses: differences $w_{\rm ex}-w_{\rm EBK}$ between 
exact quantum and EBK eigenvalues (times multiplicities).}
\label{fig7}
\end{figure}
\def\baselinestretch{2}
Using these expressions, we have calculated the first order corrections 
$\Delta w_{k,1}$ to the lowest eigenvalues of the circle billiard, by
harmonic inversion of periodic orbit signals with $s_{\rm max}=200$.
Part of the spectrum is presented in Figure \ref{fig7}.
The peak heights (squares) are the corrections 
$\Delta w_{k,1}=w_{k,1}-w_{k,0}$ times the multiplicities.
For comparison, the differences between the exact and the EBK eigenvalues 
at the positions of the EBK eigenvalues are also plotted 
(see the crosses in Fig.~\ref{fig7}).
Both spectra are in excellent agreement.
The small deviations of the peak heights arise from second or higher order 
corrections to the eigenvalues.

\def\baselinestretch{1}
\begin{figure}
\vspace{6.2cm}       
\includegraphics{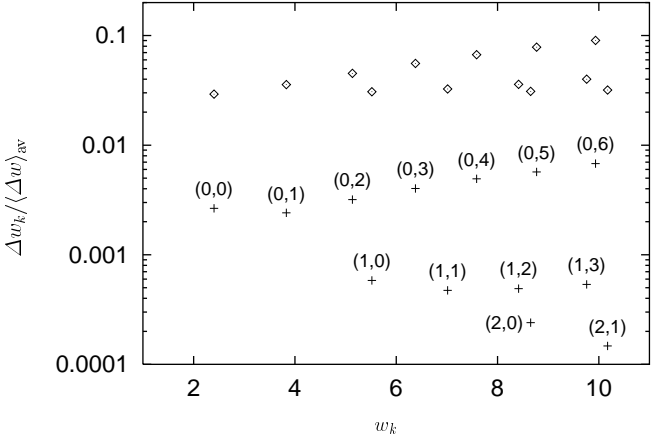}
\caption{Semiclassical errors $|w_{k,0}-w_{\rm ex}|$ (diamonds) and 
$|w_{k,1}-w_{\rm ex}|$ (crosses) of zeroth and first order approximations 
to the eigenvalues obtained by harmonic inversion, in units of the average 
level spacing $\langle\Delta w\rangle_{\rm av}\approx 4/w$. States are 
labeled by the quantum numbers $(n,m)$.}
\label{fig8}
\end{figure}
\def\baselinestretch{2}
An appropriate measure for the accuracy of semiclassical eigenvalues is 
the deviation from the exact quantum eigenvalues in units of the average
level spacings, $\langle\Delta w\rangle_{\rm av}=1/\bar\rho(w)$.
Figure \ref{fig8} presents the semiclassical error in units of the average 
level spacings $\langle\Delta w\rangle_{\rm av}\approx 4/w$ for the zeroth
order (diamonds) and first order (crosses) approximations to the eigenvalues.
In zeroth order approximation the semiclassical error for the low lying 
states is about 3 to 10 percent of the mean level spacing.
This error is reduced in the first order approximation by at least one 
order of magnitude for the least semiclassical states with radial quantum 
number $n=0$.
The accuracy of states with $n\ge 1$ is improved by two or more orders of 
magnitude.

\subsection{Reduction of required signal length via harmonic inversion 
 of cross-correlated periodic orbit sums}
\label{crosscorr}
As described in the sections above, the harmonic inversion method is able to
extract quantum mechanical eigenvalues from the semiclassical periodic orbit
sum including periodic orbits up to a {\em finite} action $s_{\rm max}$.
This means that in practice, although the periodic orbit sum does not 
converge, the eigenvalues can be obtained from a {\em finite} set of 
periodic orbits.
The required signal length for harmonic inversion depends on the mean 
density of states, i.e., $s_{\rm max}\approx 4\pi\bar\rho(w)$ 
(cf.\ (\ref{smax})).
Depending on the mean density of states, the action $s_{\rm max}$ up to 
which the periodic orbits have to be known may therefore be large.
Due to the rapid proliferation of the number of periodic orbits with 
growing action, the efficiency and practicability of the procedure depends 
sensitively on the signal length required. 
This is especially the case when the periodic orbits have to be found 
numerically.

The quantization method can be improved with the help of cross-correlated 
periodic orbit sums.
The extended response functions weighted with products of diagonal matrix 
elements discussed in Section \ref{matrixel}, in combination with the 
method for harmonic inversion of cross-correlation functions presented in
Section \ref{hi2}, can be used to significantly reduce the signal length 
required for the periodic orbit quantization \cite{Mai99c,Mai99b}.
This technique is particularly helpful for chaotic systems, where the 
periodic orbits must be found numerically and where the number of periodic 
orbits grows exponentially with their action.
However, for regular systems the number of orbits which have to be included 
can also be significantly reduced as will be demonstrated for the circle
billiard.

The basic idea is to construct a set of signals where each individual 
signal contains the same frequencies (i.e., semiclassical eigenvalues) and 
the amplitudes are correlated by obeying the restriction (\ref{restrict}).
This can be achieved with the help of the generalized periodic orbit sum 
(\ref{gab}) introduced in Section \ref{matrixel}:

A set of operators $\hat A_\alpha$, $\alpha=1,\dots,N$ is chosen. 
Following the procedure described in Section \ref{genproc}, the signals 
$C_{\alpha\alpha'}(s)$ are obtained as Fourier transform of the generalized 
response functions $g^{\rm osc}_{\alpha\alpha'}(w)$, i.e.,
\begin{eqnarray}
 g^{\rm osc}_{\alpha\alpha'}(w) &=& \sum_{\rm po}{\mathcal A}_{\rm po}
 \bar A_{\alpha,p}\bar A_{\alpha',p}e^{is_{\rm po}w} \; , \\
 C_{\alpha\alpha'}(s) &=& \sum_{\rm po}{\mathcal A}_{\rm po}
 \bar A_{\alpha,p}\bar A_{\alpha',p} \delta(s-s_{\rm po}) \; ,
\label{C_sc_corr}
\end{eqnarray}
where for integrable systems the means $\bar A_{\alpha,p}$ are defined by 
(\ref{average_reg}).
According to Sections \ref{matrixel} and \ref{genproc}, the corresponding 
quantum mechanical signal is given by
\begin{equation}
\label{C_qm_corr}
 C_{{\rm qm},\alpha\alpha'}(s) = -i\sum_k m_k 
 \langle k|\hat A_\alpha|k \rangle\langle k|\hat A_{\alpha'}|k \rangle
 e^{-isw_k} \; ,
\end{equation}
where the amplitudes have the required form (\ref{restrict}).
As in Section \ref{genproc} the semiclassical eigenvalues $w_k$ are obtained
by adjusting the periodic orbit signal (\ref{C_sc_corr}) (after convolution
with a Gaussian function) to the functional form of the cross-correlated 
quantum signal (\ref{C_qm_corr}) with the important difference that we now 
apply the extension of harmonic inversion to cross-correlation functions 
(see Section \ref{hi2}).

For the circle billiard, the mean density of states -- with all 
multiplicities taken as one -- is given by $\bar\rho(w)=w/4$.
According to (\ref{smax}), the signal length required for a single signal 
to resolve the frequencies in a given interval around $w$ is therefore 
approximately given by
\begin{equation}
 s_{\rm max} \approx 4\pi\bar\rho(w) = \pi w = 2S_H \; , 
\end{equation}
where $S_H=2\pi\bar\rho$ is the Heisenberg period (which is action instead
of time for scaling systems).
By using an $N\times N$ set of signals, it should be possible to extract
about the same number of semiclassical eigenvalues from a reduced signal 
length $s_{\rm max} \ll 2S_H$, or, vice versa, if the signal length is 
held constant, the resolution and therefore the number of converged
eigenvalues should significantly increase.

\def\baselinestretch{1}
\begin{table}
\caption{
Nearly degenerate eigenvalues of the circle billiard, 
obtained by harmonic inversion of a 
$2\times 2$ cross-correlated signal of length $s_{\rm max}=150$. 
The denotations are the same as in Table \ref{tab1}.
The nearly degenerate eigenvalues are now resolved, which for a single 
signal would have required a signal length of $s_{\rm max}\approx 500$.}
\label{tab2}
\begin{center}
\begin{tabular}{rrrrrrr}
\hline\noalign{\smallskip}
$n$ & $m$ & $w_{\rm ex}$ & $w_{\rm EBK}$ & $m_{\rm ex}$ & $w_{\rm hi}$ & $m_{\rm hi}$\\
\noalign{\smallskip}\hline\noalign{\smallskip}
1 &  4 &  11.064709   &   11.048664   &   2   &   11.048664   &    2.0665   \\ 
\smallskip
0 &  7 &  11.086370   &   11.049268   &   2   &   11.049295   &    1.9315   \\
3 &  1 &  13.323692   &   13.314197   &   2   &   13.314205   &    1.9987   \\ 
0 &  9 &  13.354300   &   13.315852   &   2   &   13.315839   &    2.0016   \\ 
\noalign{\smallskip}\hline
\end{tabular}
\end{center}
\end{table}
\def\baselinestretch{2}
To demonstrate the power of the cross-correlation technique, we first 
analyze a $2\times 2$ cross-correlated periodic orbit signal of the
circle billiard with $\hat A_{1}={\bf 1}$ the identity operator and 
$\hat A_{2}=r$. 
For comparison with the results in Section \ref{results} we choose the
same signal length $s_{\rm max}=150$.
By contrast to the eigenvalues in Table \ref{tab1} obtained from the 
one-dimensional signal the nearly degenerate states around $w\approx 11.05$
and $w\approx 13.3$ are now resolved as can be seen in Table \ref{tab2}.
Note that a signal length $s_{\rm max}\approx 500$ is required to resolve
these states without application of the cross-correlation technique.

As in all other calculations concerning cross-correlated signals, the 
results were improved by not making a sharp cut at the accumulation points 
but by damping the amplitudes of the orbits near these points.
With the same cut-off criterion at the accumulation points, the total 
number of orbits in the signal with $s_{\rm max}=150$ was about 10 times 
smaller than in the signal with $s_{\rm max}=500$ .
This means, we could reduce the required number of orbits by one order 
of magnitude.
For chaotic systems, where the number of orbits grows more rapidly 
(exponentially) with the maximum action, the improvement in the 
required number of orbits may even be better.

\def\baselinestretch{1}
\begin{table}
\caption{Maximum frequencies $w_{\rm max}$ up to which the 
spectrum could be resolved using a $N\times N$ cross-correlated signal. 
The $b_{\alpha}$ are the operators or functions of operators used to build 
the signal (see text).
$r$: distance from center, 
$L$: angular momentum in units of $\hbar w$.}
\label{tab3}
\begin{center}
\begin{tabular}{c|r|l}
\hline\noalign{\smallskip}
$N$ & $w_{\rm max}$ & $b_{\alpha}$ \\
\noalign{\smallskip}\hline\noalign{\smallskip}
1 & 45 & ${\displaystyle 1}$ \\
2 & 65 & ${\displaystyle 1,\ \langle r \rangle}$ \\
3 & 90 & ${\displaystyle 1,\ \langle r \rangle,\ L^2 }$ \\
4 & 120 & ${\displaystyle 1,\ \langle r \rangle,\ L^2,\ 
e^{-{(\langle r \rangle -0.7)^2/3}}}$ \\
5 & 130 & ${\displaystyle 1,\ \langle r \rangle,\ L^2,\ 
{1/ \langle r \rangle^2},\ e^{-{(L-1)^2/10}}}$ \\
\noalign{\smallskip}\hline
\end{tabular}
\end{center}
\end{table}
\def\baselinestretch{2}
We now investigate the number of eigenvalues which do converge for fixed 
signal length but different sets and dimension of the cross-correlation 
matrix.
Indeed, the highest eigenvalue $w_{\rm max}$ which can be resolved
increases significantly when the cross-correlation technique is applied.
However, the detailed results depend on the operators chosen.
Furthermore, with increasing dimension of the matrix, the range in which 
the transition from resolved to unresolved eigenvalues takes place becomes 
broader, and the amplitudes in this region become less well converged. 
Rough estimates of $w_{\rm max}$ for various sets of operators and fixed
signal length $s_{\rm max}=150$ are given in Table \ref{tab3}.
For some of the signals, we used the extension of the trace formula 
to functions of matrix elements, Eqs.~(\ref{gfqm}) and (\ref{gf}).
The improvement achieved by increasing the dimension of the matrix by one 
is most distinct for very small $N$; for $N\ge5$, the improvement is only 
small.
This suggests that, at given signal length and frequency range, the 
matrix dimension should not be chosen too large, i.e., there exists 
an optimal matrix dimension, which at constant signal length becomes 
larger with increasing eigenvalues $w$.

\def\baselinestretch{1}
\begin{figure}
\vspace{6.7cm}       
\includegraphics{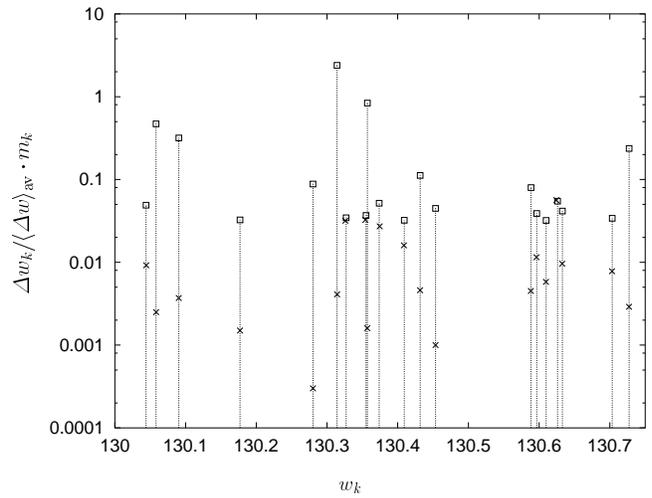}
\caption{Eigenvalues of the circle billiard in the dense part of the
spectrum obtained from a $5\times 5$ cross-correlated signal of length 
$s_{\rm max}=150=0.735S_H$ with $S_H$ the Heisenberg period.
With a single signal the required signal length would be
$s_{\rm max}\approx 2S_H$.
Squares and sticks: EBK eigenvalues. 
Crosses: results of harmonic inversion.
The peak heights give the semiclassical error $|w_{\rm EBK}-w_{\rm ex}|$ 
(squares) and the difference between harmonic inversion results and EBK 
eigenvalues $|w_{\rm hi}-w_{\rm EBK}|$ (crosses) in units of the mean 
level spacing $\langle\Delta w\rangle_{\rm av}\approx 4/w$.
The error of the harmonic inversion procedure is about an order of
magnitude smaller than the semiclassical error.}
\label{fig9}
\end{figure}
\def\baselinestretch{2}
With a $5\times 5$ signal of length $s_{\rm max}=150$, eigenvalues up to 
the region $w\approx 130$ can be resolved.
The results are presented in Figure \ref{fig9}. 
There are two points which should be emphasized:
The first point is that, even in this dense part of the spectrum, 
the error of the method is still by about one order of magnitude smaller 
than the semiclassical error, which is illustrated in Fig.~\ref{fig9} 
by the peak heights.
The squares and crosses mark the semiclassical error 
$|w_{\rm EBK}-w_{\rm ex}|$ and the numerical error 
$|w_{\rm hi}-w_{\rm EBK}|$
of the harmonic inversion procedure in units of the mean level spacing
$\langle\Delta w\rangle_{\rm av}\approx 4/w$.
The second point concerns the signal length compared to the Heisenberg 
action $S_H=2\pi\bar\rho$.
For $w=130$, one obtains $S_H\approx 204.2$. 
A one-dimensional signal would have required a signal length 
$s_{\rm max}\approx 2S_H$.
With the cross-correlation technique, we calculated the eigenvalues from a 
signal of length $s_{\rm max}=150\approx 0.735S_H$.
This is about the same signal length as required by the semiclassical
quantization method of Berry and Keating \cite{Ber90}, which, however, 
only works for ergodic systems.

In summary, our results demonstrate that by analyzing cross-correlated 
signals instead of a single signal, the required signal length can indeed 
be significantly reduced. 
Clearly, the signal length cannot be made arbitrarily small, and the 
method is restricted to small dimensions of the cross-correlation matrix. 
However, the number of orbits which have to be included can be 
very much reduced. Another advantage of the method is that not only the 
frequencies and the multiplicities but also the diagonal matrix elements
of the chosen operators are obtained by one single calculation.

\subsubsection{Including higher order $\hbar$ corrections}
In the cases discussed so far, we have constructed the cross-correlated 
signal by including different operators and making use of Eq.~(\ref{gab}). 
By this procedure, we could obtain the semiclassical 
eigenvalues from a signal of reduced length or improve the resolution of 
the spectrum at constant signal length, while simultaneously obtaining the 
diagonal matrix elements of the operators. We can now even go one step 
further and include higher $\hbar$ corrections in the signal. Here we 
make use of the results of Section \ref{hbar1}.
The first order correction term, which in Section \ref{hbar1} was 
harmonically inverted as a single signal, is now included as part of a 
cross-correlated signal. This procedure combines all the techniques 
developed in the previous sections.

Formally, the frequencies in the zeroth and first order
$\hbar$ parts of the cross-correlated signal are
not exactly the same [see the denominators in Eqs.~(\ref{gqmw}) and 
(\ref{g_int_qm})], however, as already mentioned in Section \ref{hbar1}, 
the numerically obtained values for the frequencies $\bar w_k$ 
in (\ref{Cn_qm}) are equal to the lowest order $\hbar$ eigenvalues 
rather than the exact average of the zeroth and first order eigenvalues.
In practice, the cross-correlated signal is therefore in fact
of the form  (\ref{cross_sig}).
We can now, on the one hand, improve the resolution of the spectrum, and, 
on the other hand, obtain semiclassical matrix elements and the first 
order corrections to the eigenvalues with the same high resolution by 
one single harmonic inversion of a cross-correlated signal.

\def\baselinestretch{1}
\begin{table}
\caption{Zeroth ($w_{k,0}$) and first ($w_{k,1}$) order 
semiclassical approximations to the eigenvalues of the circle billiard, 
obtained simultaneously by harmonic inversion of a $3\times 3$ 
cross-correlated signal of length $s_{\rm max}=150$. The nearly degenerate 
eigenvalues at  $w\approx 11.0$ and $w\approx 13.3$ are well resolved.}
\label{tab4}

\smallskip
\begin{center}
\begin{tabular}[t]{rrrrrr}
\hline\noalign{\smallskip}
  \multicolumn{1}{c}{$n$} &
  \multicolumn{1}{c}{$m$} &
  \multicolumn{1}{c}{$w_{\rm EBK}$} &
  \multicolumn{1}{c}{$w_{k,0}$} &
  \multicolumn{1}{c}{$w_{k,1}$} &
  \multicolumn{1}{c}{$w_{\rm ex}$} \\ 
\noalign{\smallskip}\hline\noalign{\smallskip}
   0 &   0 &    2.356194 &    2.356230 &    2.409288 &    2.404826 \\
   0 &   1 &    3.794440 &    3.794440 &    3.834267 &    3.831706 \\
   0 &   2 &    5.100386 &    5.100382 &    5.138118 &    5.135622 \\
   1 &   0 &    5.497787 &    5.497816 &    5.520550 &    5.520078 \\
   0 &   3 &    6.345186 &    6.345182 &    6.382709 &    6.380162 \\
   1 &   1 &    6.997002 &    6.997006 &    7.015881 &    7.015587 \\
   0 &   4 &    7.553060 &    7.553055 &    7.590990 &    7.588342 \\
   1 &   2 &    8.400144 &    8.400145 &    8.417503 &    8.417244 \\
   2 &   0 &    8.639380 &    8.639394 &    8.653878 &    8.653728 \\
   0 &   5 &    8.735670 &    8.735672 &    8.774213 &    8.771484 \\
   1 &   3 &    9.744628 &    9.744627 &    9.761274 &    9.761023 \\
   0 &   6 &    9.899671 &    9.899660 &    9.938954 &    9.936110 \\
   2 &   1 &   10.160928 &   10.160949 &   10.173568 &   10.173468 \\
   1 &   4 &   11.048664 &   11.048635 &   11.063791 &   11.064709 \\
   0 &   7 &   11.049268 &   11.049228 &   11.087943 &   11.086370 \\
   2 &   2 &   11.608251 &   11.608254 &   11.619919 &   11.619841 \\
   3 &   0 &   11.780972 &   11.780993 &   11.791599 &   11.791534 \\
   0 &   8 &   12.187316 &   12.187302 &   12.228037 &   12.225092 \\
   1 &   5 &   12.322723 &   12.322721 &   12.338847 &   12.338604 \\
   2 &   3 &   13.004166 &   13.004167 &   13.015272 &   13.015201 \\
   3 &   1 &   13.314197 &   13.314192 &   13.323418 &   13.323692 \\
   0 &   9 &   13.315852 &   13.315782 &   13.356645 &   13.354300 \\
   1 &   6 &   13.573465 &   13.573464 &   13.589544 &   13.589290 \\
   2 &   4 &   14.361846 &   14.361846 &   14.372606 &   14.372537 \\
   0 &  10 &   14.436391 &   14.436375 &   14.478531 &   14.475501 \\
   3 &   2 &   14.787105 &   14.787091 &   14.795970 &   14.795952 \\
   1 &   7 &   14.805435 &   14.805457 &   14.821595 &   14.821269 \\
   4 &   0 &   14.922565 &   14.922572 &   14.930938 &   14.930918 \\
   0 &  11 &   15.550089 &   15.550084 &   15.593060 &   15.589848 \\
   2 &   5 &   15.689703 &   15.689701 &   15.700239 &   15.700174 \\
   1 &   8 &   16.021889 &   16.021888 &   16.038034 &   16.037774 \\
   3 &   3 &   16.215041 &   16.215047 &   16.223499 &   16.223466 \\
   4 &   1 &   16.462981 &   16.462982 &   16.470648 &   16.470630 \\
   0 &  12 &   16.657857 &   16.657846 &   16.701442 &   16.698250 \\
   2 &   6 &   16.993489 &   16.993486 &   17.003884 &   17.003820 \\
   1 &   9 &   17.225257 &   17.225252 &   17.241482 &   17.241220 \\
   3 &   4 &   17.607830 &   17.607831 &   17.615994 &   17.615966 \\
   0 &  13 &   17.760424 &   17.760386 &   17.804708 &   17.801435 \\
   4 &   2 &   17.952638 &   17.952662 &   17.959859 &   17.959819 \\
   5 &   0 &   18.064158 &   18.064201 &   18.071125 &   18.071064 \\
\noalign{\smallskip}\hline
\end{tabular}
\end{center}
\end{table}
As an example, we built a $3\times 3$ signal for the circle billiard 
from the first order correction term given by (\ref{apo1}) and the operators 
$\hat A_1={\bf 1}$ (identity) and $\hat A_2=\hat r$. 
Again, we chose a signal length of $s_{\rm max}=150$.
By harmonic inversion of the cross-correlated signal, we obtained 
simultaneously the semiclassical eigenvalues, their first order order 
corrections, and the semiclassical matrix elements of the operator $\hat r$. 
The results for the zeroth order approximations $w_{k,0}$ to the eigenvalues
and the first order approximations $w_{k,1}=w_{k,0}+\Delta w_{k,1}$
are presented in Table \ref{tab4}.
For comparison the exact and the EBK eigenvalues are also given.
As for the results presented in Table \ref{tab2}, we were able to resolve 
the nearly degenerate states in the zeroth order approximation, which for 
a single signal would have required a signal length of 
$s_{\rm max}\approx 500$. 
Moreover, in contrast to the results of Section \ref{hbar1}, we could now 
also resolve the first order approximations to the nearly degenerate states.

\section{Conclusion}
\label{conclusion}
The harmonic inversion method has been introduced as a powerful tool for 
the calculation of quantum eigenvalues from periodic orbit sums as well 
as for the high resolution analysis of quantum spectra in terms of classical 
periodic orbits. We have demonstrated that this method, which has already 
successfully been applied to classically chaotic systems, yields excellent 
results for regular systems as well. 
Harmonic inversion has thus been shown to be a universal method, which, 
in contrast to other high resolution methods, does not depend on special 
properties of the system such as ergodicity or the existence of a symbolic 
code. 

With the harmonic inversion method, we are able to
calculate the contributions of the classical periodic orbits to the trace 
formula from the quantum eigenvalues with high precision and high resolution.
By analyzing the difference spectrum between exact and semiclassical 
eigenvalues, we could determine higher order $\hbar$ corrections to the 
periodic orbit sum of the circle billiard.

Up to now, no theory for the $\hbar$
corrections to the Berry-Tabor formula for regular systems has been
developed. We have numerically found an expression
for the first order $\hbar$ corrections to the Berry-Tabor formula by
harmonic inversion of the difference spectrum.
The same expression can be derived analytically
by using Vattay's and Rosenqvist's method for chaotic systems and
introducing some reasonable ad-hoc assumption for the circle billiard.
As this is clearly not a strict derivation, it is an interesting task for
the future to develop a general theory for the higher order $\hbar$
corrections to the trace formula for regular systems.

In addition to calculating semiclassical eigenvalues from the usual periodic 
orbit sum, we have demonstrated how further information can be extracted 
from the parameters of the classical orbits by applying the harmonic 
inversion technique to different extensions of the trace formula.
Using a generalized trace formula including an arbitrary operator, we have 
shown that the method also allows the calculation of semiclassical diagonal 
matrix elements from the parameters of the periodic orbits. 
Furthermore we have demonstrated how higher order $\hbar$ corrections to 
the semiclassical eigenvalues can be obtained by harmonic inversion of 
correction terms to the periodic orbit sums. For the case of the circle 
billiard, we found that, including the first order correction, the accuracy 
of the semiclassical eigenvalues compared to the exact quantum eigenvalues 
could be improved by one or more orders of magnitude.

Although by harmonic inversion the quantum eigenvalues can be calculated 
from a semiclassical signal of finite length, i.e., from a finite set of 
periodic orbits, the number of orbits which have to be included may still 
be large. We have demonstrated that by a generalization of the harmonic 
inversion method to cross-correlation functions the required signal length 
may be significantly reduced, even below the Heisenberg time.
Because of the rapid proliferation of periodic orbits with growing period, 
this means that the number of orbits which have to be included may be 
reduced by about one to several orders of magnitude.

\appendix
\section{Calculation of the first order $\hbar$ correction terms to 
the semiclassical trace formula}
\label{appendix}
The calculation of higher order $\hbar$ corrections to the semiclassical
eigenvalues introduced in Section \ref{hbarsec1} requires the knowledge
of the $n$th order amplitudes ${\cal A}_{\rm po}^{(n)}$ in the periodic
orbit sum (\ref{g_hbar_series}).
In this Appendix, we briefly outline the derivation of the first order
amplitudes ${\cal A}_{\rm po}^{(1)}$ and the application to the circle
billiard given in Eq.~(\ref{a1a0}).

Two different methods for the calculation of higher order $\hbar$
correction terms in chaotic systems have been derived by Gaspard and Alonso 
\cite{Alo93,Gas93} and Vattay and Rosenqvist \cite{Vat94,Vat96,Ros94}.
The latter method has been specialized to two-dimensional chaotic billiards 
in \cite{Ros94}.
Here, we follow the approach of Vattay and Rosenqvist.
However, it is important to note that both methods cannot straightforwardly
be applied to integrable systems and additional assumptios will be necessary
to derive Eq.~(\ref{a1a0}) for the circle billiard.
A general theory for the calculation of higher order $\hbar$ corrections
to the Berry-Tabor formula (\ref{BerTabForm}) for integrable systems is, 
to the best of our knowledge, still lacking.

Vattay and Rosenqvist give a quantum generalization of Gutzwiller's trace
formula based on the path integral representation of the quantum
propagator.
The basic idea of their method is to express
the global eigenvalue spectrum in terms of local eigenvalues computed in the
neighbourhood of periodic orbits.
The energy domain Green function $G(q,q',E)$ is connected to 
the spectral determinant $\Delta (E)=\Pi_n (E-E_n)$, with $E_n$ the energy 
eigenvalues or resonances, by
\begin{equation}
{\rm Tr}\,G(E)=\int dq  G(q,q,E)={d\over dE}\ln \Delta (E).
\end{equation}
The trace of the Green function can be expressed in terms of contributions
from periodic orbits
\begin{equation}
{\rm Tr}\,G(E)=\sum_{\rm p.o.}{\rm Tr}\,G_p(E),
\end{equation}
with the local traces connected to the local spectral determinants by 
\begin{equation}\label{trGp}
{\rm Tr}\,G_p(E)={d\over dE}\ln \Delta_p (E).
\end{equation}
The trace of the Green function can therefore be calculated by solving the 
local Schr\"odinger equation around each periodic orbit, which yields the
local eigenspectra.

To obtain the local eigenspectra, the ansatz
\begin{equation}
\psi=\Phi e^{iS/\hbar}
\end{equation}
is inserted into the Schr\"odinger equation, yielding the following 
differential equations for $\Phi$ and $S$.
\begin{eqnarray}
&&\partial_t S+{1\over2}(\nabla S)^2+U=0
\label{dglS}\\
&&\partial_t \Phi+\nabla \Phi \nabla S
+{1\over 2}\Phi\Delta S - {i\hbar\over2}\Delta\Phi=0,
\label{dglPhi}
\end{eqnarray}
where $U$ is the potential entering the Schr\"odinger equation.

The spectral determinant can be calculated from the local eigenvalues of the
amplitudes $\Phi$. For arbitrary energy E, the amplitudes $\Phi^l_p$ of 
the local eigenfunctions fulfill the equation
\begin{equation}\label{eigeneq}
\Phi^l_p(t+T_p)=R^l_p(E)\Phi^l_p(t),
\end{equation}
where $T_p$ is the period of the classical orbit.
Using Eq.~(\ref{trGp}), the trace formula can be expressed in terms of 
the eigenvalues $R^l_p(E)$:
\begin{eqnarray}
{\rm Tr}\,G(E)={1\over i\hbar} &&
\sum_p \sum_l \left(T_p(E)-i\hbar{d\ln R^l_p(E)\over dE}\right)\nonumber\\
&\times&\sum_{r=1}^\infty (R^l_p(E))^re^{{i\over\hbar} rS_p(E)}.
\label{qmGutzw}
\end{eqnarray}
This is the quantum generalization of Gutzwiller's trace formula and
holds exactly.

The amplitudes and their eigenvalues are now expanded in 
powers of $\hbar$:
\begin{eqnarray}
\Phi^l&=&\sum_{m=0}^\infty \left( {i\hbar\over2}\right)^m\Phi^{l(m)}\\
R^l(E)&=&\exp\left\{\sum_{m=0}^\infty \left( {i\hbar\over2}\right)^m
 C_l^{(m)}\right\}\\
&\approx& \exp( C_l^{(0)})\left(1+{i\hbar\over 2}C_l^{(1)}+\dots \right)
\label{defCl}
\end{eqnarray}
The terms $C_l^{(0)}$ yield the Gutzwiller trace formula as zeroth order
approximation, while the terms with $m>0$ give $\hbar$ corrections.

To solve Eqs.~(\ref{dglS}) and (\ref{dglPhi}) in different order of $\hbar$, 
the Schr\"odinger equation and the functions $\Phi^{l(m)}$ and $S$
are Taylor expanded around the periodic orbit,
\begin{eqnarray}
S({\bf q},t)&=&
\sum{1\over {\bf n}!}s_{\bf n}(t)({\bf q}-{\bf q}_p(t))^{\bf n}\\
\Phi^{l(m)}({\bf q},t)&=&
\sum{1\over {\bf n}!}\phi^{l(m)}_{\bf n}(t)({\bf q}-{\bf q}_p(t))^{\bf n},
\end{eqnarray}
resulting in a set
of differential equations for the different orders of the Taylor
expansions and different orders in $\hbar$. In one dimension,
these equations read explicitly: 
\begin{equation}\label{diffs}
\dot s_n-s_{n+1}\dot q 
+{1\over 2} \sum_{k=0}^n {n!\over (n-k)!k!}
s_{n-k+1}s_{k+1}+u_n=0,
\end{equation}
where $u_n$ are the coefficients of the Taylor expanded potential, and
\begin{eqnarray}
\dot\phi_n^{(m+1)}&-&\phi_{n+1}^{(m+1)}\dot q \nonumber\\
&+&\sum_{k=0}^n{n!\over (n-k)!k!}\times\nonumber\\
&&(\phi^{(m+1)}_{n-k+1}s_{k+1}
+{1\over 2}\phi^{(m+1)}_{n-k}s_{k+2})\nonumber\\
&\qquad&-\phi^{(m)}_{n+2}=0
\label{diffphi}
\end{eqnarray}
This set of differential equations can be solved iteratively.
The $l$-th eigenfunction is characterized by the condition 
$\phi_n^{(m)}\equiv 0$ for $n<l$.
The different orders of $\hbar$ are connected by the 
last term in (\ref{diffphi}).
Starting from zeroth order $\hbar$ and the lowest nonvanishing order of 
the Taylor expansion, the functions can be determined order by order.
For higher dimensional systems, the coefficient matrices obey similar
equations, and the structure of the set of equations remains the same.

To obtain the first order $\hbar$ correction to the Gutzwiller trace
formula, one has to calculate the quantities $C^{(1)}_l$.
To obtain these quantities one has to solve the set of equations
(\ref{diffphi}) up to the lowest nonvanishing first order $\hbar$
coefficient function $\phi^{l(1)}_l$, respectively.
As $\phi_n^{l(m)}\equiv 0$ for $n<l$,
this involves solving the equations for $s_2$, $s_3$ and $s_4$, and for
the zeroth order $\hbar$ coefficient functions
$\phi^{l(0)}_l$, $\phi^{l(0)}_{l+1}$ and $\phi^{l(0)}_{l+2}$.
If the initial conditions are set to be $\Phi_l^{l(0)}(0)=1$ and
$\Phi_l^{l(m)}(0)=0$ for $m>0$, the correction term $C^{(1)}_l$ is then given
by the relation 
\begin{equation}
C^{(1)}_l={\phi^{l(1)}_l(T_p)\over \exp(C^{(0)}_l)},
\end{equation}
which follows from the $\hbar$ expansion of the eigenequation (\ref{eigeneq}).

An explicit recipe for the calculation of the first $\hbar$ correction
for two-dimensional chaotic billiards is given in \cite{Ros94}.
For billiards, the potential $U$ in the Schr\"odinger equation equals
zero between two bounces at the hard wall. 
The functions $S$ and $\Phi$ now have to be Taylor expanded in two dimensions:
\begin{eqnarray}
S(x,y,t)&=&S_0+S_x\Delta x+ S_y\Delta y\nonumber\\
&+& {1\over 2}(S_{x^2}(\Delta x)^2+2S_{xy}\Delta x\Delta y
+S_{y^2}(\Delta y)^2)\nonumber \\
&+&\dots
\end{eqnarray}
and similarly for $\Phi$.
If the coordinate system is chosen in
such a way that $x$ is along the periodic orbit and $y$ is perpendicular
to the orbit,
derivatives with respect to $x$ can be expressed in terms of the derivatives
with respect to $y$ using the stationarity conditions
\begin{equation}
S_{x^{n+1}y^m}={\dot S_{x^n y^m}\over S_x},\qquad
\phi_{x^{n+1}y^m}={\dot \phi_{x^n y^m}\over S_x}.
\end{equation}
The quantity $S_x$ is equal to the classical momentum of the particle.

For the free motion between the collisions with the wall, the set of
differential equations correspoding to (\ref{diffs}) and (\ref{diffphi})
then reduces to a set of equations involving only derivatives with respect
to $y$. 
These equations can be solved analytically, with the general solution still 
containing free parameters.  Setting $S_x=1$, the first coefficient functions 
of the Taylor expanded phase are given by:
\begin{eqnarray}
S_{yy}(t)&=&{1\over t+t_0}\label{gen_sol_s1}\\
S_{yyy}(t)&=&{A\over(t+t_0)^3}\\
S_{yyyy}(t)&=&-{3\over(t+t_0)^3}+{B\over(t+t_0)^4}+{3A^2\over(t+t_0)^5}
\end{eqnarray}
where $t_0$, $A$ and $B$ are free parameters. 
For given $l$, the  first nonvanishing coefficients of the amplitude read
\begin{eqnarray}
\phi^{(0)}_{y^l}(t)&=&E\left({t_0\over t+t_0}\right)^{l+1/2}\\
\phi^{(0)}_{y^{l+1}}(t)&=&{E\over (t+t_0)^{l+3/2}}
\left[C+(l+1)^2{A\over2} {t_0^{l+1/2}\over(t+t_0)}\right]\\
\phi^{(0)}_{y^{l+2}}(t)&=&{E\over (t+t_0)^{l+5/2}}
\Biggl\{D+    {1\over t+t_0} \times\nonumber\\
&&\Bigl[(l+2)^2{AC\over 2} + 
      (l+2)(l+1)({l\over 3}+{1\over 2}) {B\over 2} t_0^{l+1/2}\Bigr]
  \nonumber\\
&+&
{A^2 t_0^{l+1/2}\over 2(t+t_0)^2}\times\nonumber\\
&&\Bigl[{1\over 4}(l+2)^2(l+1)^2+
       {3\over 2}(l+2)(l+1)({l\over 3}+{1\over 2})\Bigr] 
\Biggr\}\nonumber\\
\label{gen_sol_phi3}
\end{eqnarray}
Again, $D$ and $E$ are free parameters.

At the collisions with the hard wall, 
the phase and amplitude have to obey the bouncing
conditions
\begin{eqnarray}
S(x,y,t_{-0})&=&S(x,y,t_{+0})+i\pi
\label{bounceS}\\
\Phi(x,y,t_{-0})&=&\Phi(x,y,t_{+0}),
\label{bouncePhi}
\end{eqnarray}
from which the bouncing conditions for the coefficients of
the Taylor expanded functions $S$ and $\Phi$ can be derived.
While the general solutions between the bounces are valid for all
billiards, the bouncing conditions in their Taylor expanded form
depend explicitely on the shape of the hard wall.

An additional condition which the solutions $S$ and $\Phi$ have to obey
is periodicity along the orbit. With every traversal the phase gains
the same constant contributions at the collisions with the wall. The
derivatives of the phase are periodic. The amplitude collects the same
factor with each traversal, which means that all Taylor coefficients
of the amplitude are periodic apart from a constant factor.
These conditions together with the bouncing conditions determine
the values of the free constants in the general solutions between the
collisions.

The solutions can in general be found numerically, by choosing suitable
initial conditions and following the evolution of the phase and amplitude
functions along the orbit. After several iterations around the orbit,
the parameters should converge against their periodic solution.
The correction terms $C_l^{(1)}$ are then given by the integral
\begin{equation}\label{cl1_int}
C_l^{(1)} =
\int{\phi_{y^{l+2}}^{l(0)}+\phi_{y^lx^2}^{l(0)}\over \phi^{l(0)}_{y^l}}dt,
\end{equation}
which can be computed explicitly from the solutions found above.

As already explained, this method is designed for chaotic systems,
as its derivation is based on the assumption that the periodic orbits
are isolated.
Nevertheless, we obtain reasonable results when applying
the method to the circle billiard, 
taking one periodic orbit from each rational torus and introducing
some additional assumptions.

Because of the symmetry of the orbits, we can assume that every side
of the orbit contributes in the same manner to the $\hbar$ correction term
for the whole orbit. This means, if we reset $t=0$ at the
start of each side, the free parameters in the general solutions 
(\ref{gen_sol_s1}) to (\ref{gen_sol_phi3}) must
be the same for each side, apart from the parameter $E$,
which collects the same factor during every collison with the wall.
With these assumptions, the differential equations can be solved
analytically. 

However, it turns out that the bouncing conditions resulting from 
(\ref{bounceS}) and (\ref{bouncePhi})
are not sufficient to determine all free parameters, as some of
the conditions are automatically fulfilled. 
We need additional conditions for the parameters.
These can be obtained from the rotational symmetry of the system:
Because of this symmetry, we can assume that
the amplitude of the wave function does not depend on the polar angle 
$\varphi$. The same holds for all derivatives of the amplitude with 
respect to the radius $r$. 
For the zeroth order $\hbar$ amplitudes, expressed in
polar coordinates $(r,\varphi)$, this gives us the additional conditions
we need:
\begin{equation}
{\partial\over \partial\varphi}
{\partial^n\Phi^{l(0)}\over\partial r^n}=0.
\end{equation}
If we further assume that the phase separates in polar coordinates
\begin{equation}\label{sep_s}
S(r,\varphi)=S_r(r)+S_\varphi(\varphi), 
\end{equation}
which implies that all mixed derivatives vanish, it turns out that
we do not need the bouncing conditions
at all. All parameters can be determined from the symmetry of the system,
and the bouncing conditions are automatically fulfilled.
We considered only the case $l=0$, for which we used
Eq.~(\ref{sep_s}) together with the conditions 
\begin{equation} 
{\partial\over \partial\varphi}\Phi^{0(0)}=0,
\qquad {\partial\over \partial\varphi}
{\partial\Phi^{0(0)}\over\partial r}=0.
\end{equation}
Our final results for the constant parameters in 
Eqs.~(\ref{gen_sol_s1}) to (\ref{gen_sol_phi3}) are
\begin{eqnarray}
t_0&=&-\sin\gamma\\
A&=& -\cos\gamma\\
B&=&0\\
C&=&0\\
D&=&-{i\over 2}\sin^{1/2}\gamma
\end{eqnarray}
with $\gamma$ as defined in Section \ref{hbarsec1} (see Fig.~\ref{fig1}).
The radius of the billiard was taken to be $R=1$. 
Inserting these solutions in (\ref{cl1_int}) finally leads to
\begin{equation}
C^{(1)}_0=
M_r\left({1\over 3\sin\gamma}-
         {5\over 6\sin^3\gamma}\right)
\end{equation}
where $M_r$ is the number of sides of the orbit.

The first order amplitudes ${\mathcal A}^{(1)}$
are obtained by inserting the $\hbar$ expansion
(\ref{defCl}) in the trace formula (\ref{qmGutzw}) and comparing the result
with the $\hbar$ expansion (\ref{g_hbar_series}).
In the units we have used here (radius $R=1$ and momentum
$\hbar k=1$), the scaling parameter $w$ is equal to $\hbar$.
If we use only the $l=0$ contributions and assume that the terms
$\exp(C^{(0)}_0)$ are equal to the amplitudes given by the Berry-Tabor
formula, we finally end up with the expression (\ref{a1a0}).
Although we cannot strictly justify the last step, our analysis of 
the quantum spectrum in Section \ref{hbar2} provides strong numerical 
evidence that Eq.~(\ref{a1a0}) is correct.
It will be an interesting task for the future to develop a general theory
for the $\hbar$ correction terms of integrable systems and thus to provide 
a more rigorous mathematical proof of Eq.~(\ref{a1a0}).

\section*{Acknowledgements}
This work was supported by the Deutsche Forschungsgemeinschaft.

\end{document}